%% file: main.tex
\documentclass[12pt]{article}
\usepackage{graphicx} 

\usepackage{xcolor}
\usepackage{multirow}
\usepackage{subcaption}
\usepackage[obeyspaces,spaces]{xurl}

\usepackage{longtable}

\usepackage[natbibapa]{apacite}
\usepackage{array}
\newcolumntype{C}[1]{>{\centering\arraybackslash}p{#1}}
\newcolumntype{P}[1]{>{\raggedright\arraybackslash}p{#1}}

\providecommand{\keywords}[1]{\textbf{Keywords:} #1}

\usepackage{mathptmx}

\usepackage[margin=1.5in]{geometry}

\usepackage{authblk}

\title{Ireland in 2057: Projections using a Geographically Diverse Dynamic Microsimulation}

\author[1]{Seán Caulfield Curley}
\author[1]{Karl Mason}
\author[1]{Patrick Mannion}
\affil[1]{School of Computer Science, University of Galway}

\begin{document}

\maketitle

\begin{abstract}

    This paper presents a dynamic microsimulation model developed for Ireland, designed to simulate key demographic processes and individual life-course transitions from 2022 to 2057. The model captures four primary events: births, deaths, internal migration, and international migration, enabling a comprehensive examination of population dynamics over time. Each individual in the simulation is defined by seven core attributes: age, sex, marital status, citizenship, whether the person was living in Ireland in the previous year, highest level of education attained, and economic status. These characteristics evolve stochastically based on transition probabilities derived from empirical data from the Irish context. 
Individuals are spatially disaggregated at the Electoral Division level. By modelling individuals at this granular level, the simulation facilitates in-depth local analysis of demographic shifts and socioeconomic outcomes under varying scenarios and policy assumptions. The model thus serves as a versatile tool for both academic inquiry and evidence-based policy development, offering projections that can inform long-term planning and strategic decision-making through 2057. The microsimulation achieves a close match in population size and makeup in all scenarios when compared to Demographic Component Methods. Education levels are projected to increase significantly, with nearly 70\% of young people projected to attain a third level degree at some point in their lifetime. The unemployment rate is also projected to decrease as a result of the increased education levels.
\end{abstract}


\section{Introduction}
\label{sec:intro}

Microsimulation models are powerful tools for analysing and projecting demographic and socioeconomic changes at the individual level. By simulating life events such as births, deaths, and migration these models provide nuanced insights into the dynamics that shape populations over time.
In 1957, \citet{orcutt1957firstMicrosim} outlined the limitations of macro-scale economic systems, and how micro-scale systems addressed many of those challenges, Since then, microsimulations have been deployed to tackle a wide range of issues including health \citep{wu2022GB, may2022healthIreland}, income \citep{odonoghue2021incomeCovid, emmenegger2024IncomeGermany} and demography \citep{demosimCanada2025, munnich2021Germany}. In recent times, there has been a particular increase in the number of papers being published in the dynamic microsimulation sub-field \citep{odonoghue2025survey}. Dynamic microsimulations involve simulating individual units over time to analyse the temporal effect on the topic in question. 

As outlined by \citet{odonoghue2025survey}, they can be distinguished by three broad modelling choices: discrete or continuous time, aligned or non-aligned and open or closed systems. DYNASIM \citep{urban2016dynasim}, DEMOSIM \citep{demosimCanada2025}, and MikroSim \citep{munnich2021Germany} are closed, aligned and discrete, while models such as LifePaths \citep{spielauer2013Lifepaths} are open, non-aligned and continuous. There are strengths and limitations of each approach but detailed analysis of these is beyond the scope of this paper. However, there is rarely an equal amount of data available that is suitable for both approaches. Thus, the type of model implemented is largely decided by what type of model could leverage the most data (and therefore be both most complex and most realistic). However, as noted by \citet{vanImhoff1998specificationRandomness}, it must be mentioned for each explanatory variable added to the model, the degree of randomness of the outputs grows.  Many of the previous dynamic simulation models have been developed by national institutions like Statistics Canada \citep{demosimCanada2025} or the Urban Institute \citep{urban2016dynasim} and are not publicly available. There has been a push in recent times towards open source models to allow users to modify code and simulation scenarios for their purposes \citep{bronka2025SimPaths}. 

Traditionally, population projections, including those in Ireland, have been carried out using Demographic Component Method (DCM) \citep{Smith2013DCM}. DCM methods are computationally relatively simple and do not require a significant amount of data. However, their simplicity is also a downside because it means complex dynamic behaviours cannot be accounted for. Similarly, it is not possible to model interactions between individuals. DCM models are also more suited to projections for countries and large regions than small areas because of factors such as poor quality data, changes in small area borders, ``empty cells'', etc. It is unsurprising then, that without widely accepted methods for small area projections, national and regional projections have received much more attention, as noted by \citet{mazzuco2020developments}. However, attempts to rectify this research gap have been made recently. In their review article, \citet{wilson2022methodsForSmallAreas} summarised the state-of-the-art in small area population forecasts and highlighted how few attempts at using microsimulation there were at the time of publishing in 2021.

One example of small area projections using microsimulation is the model proposed by \citet{marois2015montreal} of Montreal's municipalities' populations. In that paper, local contextual factors such as housing stock were accounted for when considering how the population may evolve over time. \citet{lomax2022UK} proposed an open-source model for population projections in small areas using microsimulation with a case study of the ``Arc'' region of the UK. Their study also included an aspect on land-use change, and so results were presented for four different scenarios concerned with the rate of new dwelling completions. Finally, Norway's National Statistics Agency recently published a paper in which they compare future small area (municipality) populations generated by microsimulation to their existing DCM projected populations \citep{jia2023norway}. There is a strong overlap between the aggregate results which is promising considering that a microsimulation achieving the same aggregate results as a DCM will also contain much richer output information than the DCM population.

One prominent topic of research in the study of small areas is the stability of small areas parameters. For example, a study of the fertility schedules in Norway showed how using ``rates'' when small populations are under investigation can lead to large relative fluctuations in statistics year-to-year \citep{leknes2021norwaySmallAreas}. They give the example of the municipality with the median population size which has 27 women aged 30. If the true fertility rate is 0.11, the expected number of children would be 3. However, a naive random draw would simulate exactly 3 births in the area in less than a quarter of cases. There will also be small be non-zero probabilities that there are 0, or 6 births. \citet{leknes2021norwaySmallAreas} tackle this problem by employing a hierarchical Empirical Bayes strategy. This strategy leverages the fact that municipalities are nestled within regions, which are themselves nestled within the country to weigh local estimates according to their corresponding regional and national estimates. The ``reliability'' of both the local and regional estimates can be calculated which is based on the area's population as well as the variation in the target statistic amongst all of the region's municipalities. Biasing draws in this way allows for much more stable results than would be achieved from random draws.

Migration can also be modelled using Bayesian approaches. \citet{zens2025bayesianMigration} proposed a Bayesian model to estimate counts in potentially small demographic subpopulations with an application to immigration flows in Austria. These immigration flows are reported by each immigrant's year of age, sex and country of origin. Clearly, there is the potential for very small populations with these tight constraints. The proposed approach includes numerous indicators for the country of origin including EU membership, gross national income per capita and number of battle deaths as covariates. Smoothing across the age dimension is also performed. The model shows excellent performance on the Austrian case study. Finally, mortality can also be modelled using Bayesian frameworks. \citet{alexander2017mortalityBayesian} outlined a Bayesian model which pooled information across space (considered its neighbours' results) to estimate mortality rates for counties in the USA as well as \textit{departements} in France. As in the fertility case, the fewer data available in an area, the closer its parameters would be estimated towards the mean. On the other hand, a large area would not borrow much information from the mean values. Therefore, there would be higher confidence that a large area with atypical parameters is truly atypical compared to a small area.

This paper follows the trend towards open-sourcing models by presenting a comprehensive open-source dynamic microsimulation model developed for Ireland. Ireland has been fortunate to be the subject of a number of microsimulation and agent-based modelling (ABM) studies in recent times.  The Economic and Social Research Institute (ESRI) used survey and register data to model Ireland's tax-benefit system \citep{keane2023esriTAX}. The concern of their paper is related primarily to the economy, and so there is not a focus on demographic change. \citet{may2022healthIreland} did model changes to Ireland's population in the future, however they were concerned with the health and use of services among older people in Ireland. \citet{odonoghue2013SMILE} pioneered microsimulation in Ireland by creating a model similar to this one which creates a synthetic population for Ireland described by many of the same characteristics as this model. However, their model was static in nature (it did not model population change) and closed-source. 

On the ABM side, there have been a number of papers published which treat individuals as agents and use simulation to model the spread and impact of COVID-19 in Ireland \citep{hunter2021covidCounty, HUNTER2023covidContactTracing}. 
One goal of this paper is that by publishing a method of generating a representative population for Ireland for future years, similar ABMs to those simulating the spread of COVID could be utilised to analyse the impacts of a future epidemic before it happens. 
Policies to mitigate the effects of such an epidemic could also be simulated. There are also a number of Irish ABMs dealing with renewable energy \citep{faiud2025karlsDairyPaper, MELES2022heatPumps}. Topics such as the adoption of solar panels or renewable home heating are highly dependent on the nature of the adopter. Therefore, a spatially diverse representative dataset could aid in realistically modelling the future uptake of renewable energy concepts. 

A large contributing factor to the breadth of research performed to date in Ireland is the wealth of data provided by the Central Statistics Office (CSO). Data featured on the \url{data.cso.ie} repository includes Census data, Small Area Population Statistics (SAPS), and the results of the CSO's population projections. The CSO publish a report a few years after each Census outlining the results of their population projections \citep{cso2023projections} which are carried out using the DCM \citep{Smith2013DCM}. The DCM method is a widely used and well accepted method in demography, and so the results of this method are very trustworthy. Their most recent report projects the Irish population from 2023 to 2057. However, the code to replicate the CSO's results is not available, so researchers cannot test their own simulation scenarios. Furthermore, DCM is conducted at the macro (NUTS3 region\footnote{The Nomenclature of Territorial Units for Statistics (NUTS) are a standard for referencing country subdivisions. In Ireland, the NUTS3 regions are Border, West, Mid-West, Mid-East, South-East, South-West, Dublin and Midland}) rather than micro scale, so fine-grained analysis is not possible. 


In this paper, we propose the Socio-Economic Model for Irish Projections (SEMIPro). Similarly to the CSO's model, SEMIPro simulates four fundamental life events: births, deaths, internal migration, and international migration. Each individual in the simulation is characterized by a set of core attributes: age, sex, marital status, citizenship, whether the person lived in Ireland in the year prior, highest level of education attained, and economic status. These characteristics evolve stochastically over time, influenced by event-specific transition probabilities that are informed by empirical data and theoretical assumptions tailored to the Irish context.

In terms of the geographical diversity of demographic behaviour in Ireland, there is significant heterogeneity between behaviours within counties in the same regions, and also amongst Electoral Divisions (EDs - the smallest legally defined administrative areas in Ireland) within the same county. To use County Galway as an example, the ED of Knockboy in Connemara had 27 people enumerated in the 2022 Census with a usual residence one year ago outside of Ireland. Knockboy's neighbouring ED, Skannive had just 3, despite having a population approximately 70\% of the size of Knockboy. In terms of recent immigrants from a different county in Ireland, only 2 such people were enumerated in Slievaneena (also in Connemara) while 23 were enumerated in its neighbour, Furbogh. This is despite Slievaneena's population being approximately half of Furbogh's, a proportion significantly higher than the approximate tenth of Furbogh's different-county immigrants.


\begin{table}[!htb]
    \centering
    \begin{tabular}{|C{0.08\textwidth}|C{0.08\textwidth}|C{0.08\textwidth}|C{0.06\textwidth}|C{0.06\textwidth}|C{0.1\textwidth}|C{0.08\textwidth}|C{0.08\textwidth}|C{0.07\textwidth}|}
\hline
    \textbf{Sex} & \textbf{Border}  & \textbf{Dublin}  & \textbf{Mid-East}  & \textbf{Mid-West}  & \textbf{Midland} & \textbf{South-East} & \textbf{South-West} & \textbf{West} \\
    \hline
    Female & 3.4\% & 5.8\% & 3.4\% & 5.4\% & 3.5\% & 3.6\% & 4.9\% & 5.2\% \\
    \hline
    Male & 3.1\% & 5.2\% & 3.0\% & 4.9\% & 2.7\% & 3.1\% & 4.4\% & 4.3\% \\
    \hline
\end{tabular}
    \caption{The percentage of adults aged 25-34 in each region who are students.}
    \label{tab:regionalAdultStudentRates}
\end{table}

In terms of fertility, the Total Fertility Rate (TFR – the number of children a woman is expected to have in her lifetime) in the Dublin region is almost 0.2 births lower than the next lowest fertility region, West. The West's fertility rate is approximately 0.3 births births lower than the Border, the region with the highest fertility in Ireland. Education is another area in which their are relatively large geographical differences, as shown in Table \ref{tab:regionalAdultStudentRates}. It is evident that the regions containing Ireland's traditional universities (Dublin, Mid-West, West and South-West) have higher rates of adults in education than those without a traditional university.

This microsimulation contributes to the growing body of demographic modelling literature by integrating multiple life-course events into a unified framework \citep{pohl2024statsim, jia2023norway, demosimCanada2025}. Moreover, by embedding detailed individual attributes, the model allows for a rich exploration of population heterogeneity and differential outcomes across subgroups. The simulation is designed to support both academic inquiry and policy development in Ireland, enabling users to evaluate the potential demographic consequences of various policy interventions and socioeconomic trends.
In the sections that follow, we describe the different scenarios considered for projection, the structure of the model, the data sources and validation methods used, and discuss some interesting results presented for one reference set of assumptions.

\section{Data}
\label{sec:data}
A synthetic population adapted from those generated by \citet{CAULFIELDCURLEY2025mine} using Iterative Proportional Fitting (IPF) is used to represent the base Irish population in 2022. 
In a previous paper, \citet{CAULFIELDCURLEY2025mine} generated four different synthetic populations for Ireland using Iterative Proportional Fitting (IPF), Conditional Probabilities, Simulated Annealing and Genetic Algorithms. Individuals in the synthetic populations were described by 6 characteristics: age, sex, marital status, housing size, primary economic status and highest level of education attained. Populations were generated by matching 2022 Labour Force Survey (LFS) microdata to CSO SAPS aggregate statistics. Populations were generated at the ED level. EDs are the smallest legally defined administrative area in Ireland, and thus provide fine grained results. IPF was chosen as the static generation method for this paper as it was  found to significantly outperform the other methods in achieving close matches to the target ED's marginals. 
IPF is also widespread in the literature and has been used to generate synthetic populations for places like Sweden \citep{tozluoglu2023sweden} and Canada \citep{Predhumeau2023Canada}.

 One of the goodness-of-fit tests used by \citet{CAULFIELDCURLEY2025mine} is the number of Non-Fitting Tables (NFTs) in an ED. NFTs are calculated by performing a $\chi^2$ similarity test, where all of the proportions for a given characteristic in an ED e.g., the proportions of the population with each of the possible marital statuses, are compared. If the sampling distribution of the ED's actual population and the synthetic population is determined to be statistically significant (sampled from a different source), then we say that the synthetic population did not achieve a good fit for that characteristic and denote it as an NFT. This is repeated for each of the 6 characteristics that are modelled in the study. Therefore, a synthetic population ``perfectly'' representing its corresponding ED's actual population would have 0 NFTs out of 6.


\begin{table}[!htb]
    \centering
    \begin{tabular}{|C{0.25\linewidth}|C{0.475\linewidth}|C{0.15\linewidth}|}
    \hline
    \textbf{Characteristic} & \textbf{Possible Values} & \textbf{Value in Source Code}\\
    \hline
    \multirow{1}{\linewidth}{Age} & 0-105 & Relevant Integer\\
    \hline
    \multirow{2}{\linewidth}{Sex} & Female & F \\
    & Male & M \\
    \hline
    \multirow{4}{\linewidth}{Marital Status} & Married & MAR \\
    & Single & SGL \\
    & Separated & SEP \\
    & Widowed & WID \\
    \hline
    \multirow{4}{\linewidth}{Citizenship} & Ireland & IE \\
    & UK & UK \\
    & EU27 excluding Ireland & EU \\
    & Rest of World & RW \\
    \hline
    \multirow{1}{\linewidth}{Moved to Ireland in last year} & True or False & Relevant Boolean \\
    \hline
    \multirow{8}{\linewidth}{Highest Level of Education Attained} & No Formal Education & NF \\
    & Primary Education & P \\
    & Lower Secondary & LS \\
    & Upper Secondary & US \\
    & Post-Leaving Certificate & PLC \\
    & Higher Certificate & HC \\
    & Undergraduate Degree & DEG \\
    & Postgraduate Degree & PD \\
    & Doctorate & D \\
    \hline
    \multirow{8}{\linewidth}{Primary Economic Status} & At Work & W \\
    & Student & S \\
    & Looking After Home/Family & LAHF \\
    & Retired & R \\
    & Unable to Work Due to Permanent Sickness or Disability & UTWSD \\
    & Other & OTH \\
    & Unemployed & UNE \\
    & Not Applicable (for children) & NA \\
    \hline
    \end{tabular}
    \caption{The characteristics used to describe individuals along with their values in the source code}
    \label{tab:characteristics}
\end{table}

Table \ref{tab:characteristics} outlines all of the possible characteristic values that an individual in the microsimulation may have. The maximum age of an individual in the simluation is 105 years old due to a lack of data beyond that point. Although separated and divorced people are enumerated distinctly in the Census, they are combined in the LFS, so the microsimulation also bands them together under the same characteristic value. The education levels shown in the table are from Ireland's National Framework of Qualifications (NFQ) levels which is how education is enumerated in the Census. However, the LFS uses the ISCED framework to track educational attainment. An approximate match between the levels is achieved by converting the characteristic of those with an ISCED level 4 qualification to a new label of ``Post Leaving Certificate'' and those with an ISCED level 6 qualification to a new label of ``Undergraduate Degree''. Types of unemployment (looking for first regular job, short- and long-term unemployed) are also counted individually in the SAPS but given under the broad header of ``Unemployed'' in the LFS. Similarly to the ``Separated'' case, here the more detailed SAPS descriptions are summed and given the more general description of ``Unemployed''. 

\begin{table}[!htb]
    \centering
    \begin{tabular}{|c|c|}
    \hline
    \textbf{SEMIPro Values} & \textbf{LFS/SAPS values} \\
    \hline
    \multirow{2}{*}{Separated (Marital Status)} & Separated \\
    & Divorced \\
    \hline
    \multirow{3}{*}{Unemployed (Economic Status)} & Looking for First Full-Time Job \\
    & Short-Term Unemployed \\
    & Long-Term Unemployed \\
    \hline
    Primary (Education) & ISCED 1 \\
    \hline
    Lower Secondary (Education) & ISCED 2 \\
    \hline
    Upper Secondary (Education) & ISCED 3 \\
    \hline
    Post-Leaving Cert (Education) & ISCED 4 \\
    \hline
    Higher Certificate (Education) & ISCED 5 \\
    \hline
    Degree (Education) & ISCED 6 \\
    \hline
    Postgraduate Diploma (Education) & ISCED 7 \\
    \hline
    Doctorate (Education) & ISCED 8 \\
    \hline
    \multirow{2}{*}{EU (Citizenship)} & EU15 (Excluding Ireland and UK) \\
    & EU (Excluding Ireland, UK, and EU27) \\
    \hline
    \multirow{2}{*}{Rest of World (Citizenship)} & North America \\
    & Rest of World \\
    \hline
    \end{tabular}
    \caption{Mapping between SEMIPro characteristic values and the LFS/SAPS characteristics which were grouped or renamed to achieve equivalence}
    \label{tab:characteristic_mapping}
\end{table}

In the previous study by \citet{CAULFIELDCURLEY2025mine}, citizenship and previous residence one year ago were not included as characteristics. These characteristics were added in this study to achieve more detailed international migration flows as well as more realistic characteristics of international immigrants. The mapping between SEMIPro and LFS/SAPS characteristic values is outlined for these characteristics as well as those mentioned in the previous paragraph in Table \ref{tab:characteristic_mapping}. Housing size was not included because individuals are not grouped into households in this study. Another alteration made to the IPF generation procedure used by \citet{CAULFIELDCURLEY2025mine} is that the ``empty cell'' problem is tackled by adding a small weight ($1\mathrm{e}{-4}$) to all valid combinations of characteristics. This is in keeping with previous evaluations of the performance of IPF \citep{lovelace2015ipfSmallWeight} as Iterative Proportional Updating \citep{ye2009IPU}. Care was taken to avoid allocating this weight to invalid combinations of characteristics such as children with a marital status other than ``Single'' or a primary economic status other than ``NA''.

The IPF approach used here achieves a ``perfect'' match for 3,413 of the 3,417 EDs utilising the IPFP function from the MIPFP R package \citep{barthelemy2018mipfp}. With regards to the 4 EDs who have an NFT, each of them narrowly fail the $\chi^2$ test for the 'Moved to Ireland in last year' characteristic. All 4 of the failing EDs are amongst the top 12 most populated EDs in Ireland (and therefore have some of the highest recent immigrant counts also). It has been shown in the literature that there are differences in socioeconomic characteristics and outcomes for migrants and non-migrants in Ireland \citep{mcginnity2025esriIntegration}. Therefore, one reason why IPF may be struggling to match these EDs correctly could be the relatively small number of recent immigrants in the microdata not adequately representing the distributions of characteristics amongst the failing EDs' recent immigrants. In terms of the operation of the microsimulation, the role of the ``Moved to Ireland in last year'' characteristic is entirely facilitative. As explained in Section \ref{sec:international_migration}, the characteristic is used in the accurate sampling of characteristics for international immigrants. It is not explicitly used as an weighting factor in modules like education, employment, etc., due to a lack of data and the expectation that the same effects will be applied implicitly by characteristics like citizenship, age, etc. Therefore, we leave perfect matching of these 4 EDs to future work.

\section{Methods}
\label{sec:methods}
The entire microsimulation is implemented in Python and makes use of the Object-Oriented Programming paradigm to ensure the code is readable and extendable. In order to ensure that those who are not experienced in Python can easily read and run the code, only base Python and common libraries such as Pandas, NumPy and tqdm (for progress tracking) are used. Each run of the microsimulation takes approximately 5 hours on a laptop with 32GB of RAM and and Intel i7-1165G7 CPU. The output of a microsimulation is a list of CSVs, one for each ED, logging the ED's starting population, number of births, number of deaths, net internal migration, net international migration and ending population. The final population is also stored as a list of individuals in a pickle file (Python's native object serialiser). The user can also specify if populations should be stored at checkpoints during the simulation e.g., every 5 years, for more fine-grained analysis. The choropleths included in the following results were generated using Plotly Python package. The code is available at \url{https://github.com/SCC-git/ireland-microsim}.

\input{regions_and_counties}

The CSO's regional population projections are used both for validation and alignment. Those projections are performed at the NUTS3 region level. The list of Ireland's NUTS3 regions along with their constituent counties is given in Table \ref{tab:regionsCounties}. Any reference to ``region'' for the remainder of this paper refers to these NUTS3 regions.

\subsection{Simulation Scenarios}
\label{sec:scenarios}
There are 3 possible assumptions for international migration which are the same as those given in the CSO’s population projections and are denoted as M1, M2, and M3. The magnitude of migration for each assumption are as follows:
\begin{itemize}
    \item \textbf{M1}: Net migration starting at +75,000 and incrementally decreasing to +45,000 per annum by 2027. Net migration remaining at this level from 2027 on.
    \item \textbf{M2}: Net migration starting at +75,000 and incrementally decreasing to +30,000 per annum by 2032. Net migration remaining at this level from 2032 on.
    \item \textbf{M3}: Net migration starting at +75,000 and incrementally decreasing to +25,000 per annum by 2027 and +10,000 by 2032. Net migration remaining at this level from 2032 on.
\end{itemize}

As in the CSO's regional population projections, 1 internal migration scenario is modelled. The county-to-county migration flows for this scenario are retrieved from the average of the 2016 and 2022 Census values of county’s populations categorised by usual residence one year prior.

\subsection{Population Initialisation - Marriage}
\label{sec:initialisation_marriage}
The previous section outlined how a representative static population is generated. However, there are still a number of steps that must be taken to convert from this static population to one that is suitable for dynamic microsimulation. Partnerships are formed by assigning individuals with the marital status of ``MAR'' a spouse also living in their current ED unless no more married individuals are available, in which case they are assigned a spouse living in the same NUTS3 region as them. Any married people without spouses after this step (from having unequal numbers of married men and women) are assumed to be married to an individual living outside of Ireland. The level of dissimilarity between a married person and each of their potential partners is calculated as the distance between the potential couple’s ages and education level. This approach is adapted from the approaches taken in MikroSim \citep{munnich2021Germany} and DYNASIM \citep{urban2016dynasim} where mates are matched on age, race and education. A person’s spouse is then chosen by sampling the top 20 least dissimilar available partners according to a Dirichlet distribution. Three percent of marriages are assumed to be same sex and there is an approximately equal number of same sex marriages between males and females. This is in keeping with the current rates of same sex marriage in Ireland. 

\subsection{Population Initialisation - Education}
\label{sec:initialisation_education}

Any adults from the static population with an education of ``NA'' (Not Applicable) or ``NS'' (Not Stated) are assigned a valid education by sampling randomly from the distribution of education levels for people of the same age group and sex. The static population generation method gives all individuals with the economic status of ``S'' (Student) a highest level of education of ``NA'', in accordance with the LFS. The education module in this simulation necessitates knowing the highest level of education achieved for all those older than primary schoolchildren. Therefore, education levels are estimated for all students. The age of individuals under the age of 18 is used to estimate their class (the Irish equivalent to grade). It is then assumed that they have achieved the highest level of education up to that point e.g. a person studying for an Upper Secondary qualification is assumed to have achieved a Lower Secondary qualification. 

Young adult students (between 18 and 25) are assumed to be studying for one of the education levels above Upper Secondary and are assigned to their prospective level of education based on the proportion of students in each course in 2022. Students with a prospective education of ``PLC'' (Post Leaving Certificate) or ``HC'' (Higher Certificate) are assigned a highest level of education achieved of US. Students with a prospective education of an Undergraduate Degree (DGR) are randomly assigned a highest level of education achieved from the choices US, PLC, and HC according to the enrolment rates for those courses. Students with a postgraduate prospective level of education (postgraduate diploma/masters or doctorate) are assigned the highest level of education below their highest level of education achieved. 

All students are randomly assigned a graduation date in the future based on the estimated duration of their prospective highest level of education. Course durations are based on empirical data. Students over the age of 25 are assigned a highest level of education achieved according to the education proportions among the non-student adult population. Their prospective highest level of education is estimated based on the enrolment rates for adults with their highest level of education achieved to date. 

\subsection{Ordering of Modules}
\label{sec:ordering_of_modules}



\begin{figure}
    \centering
    \includegraphics[width=\linewidth]{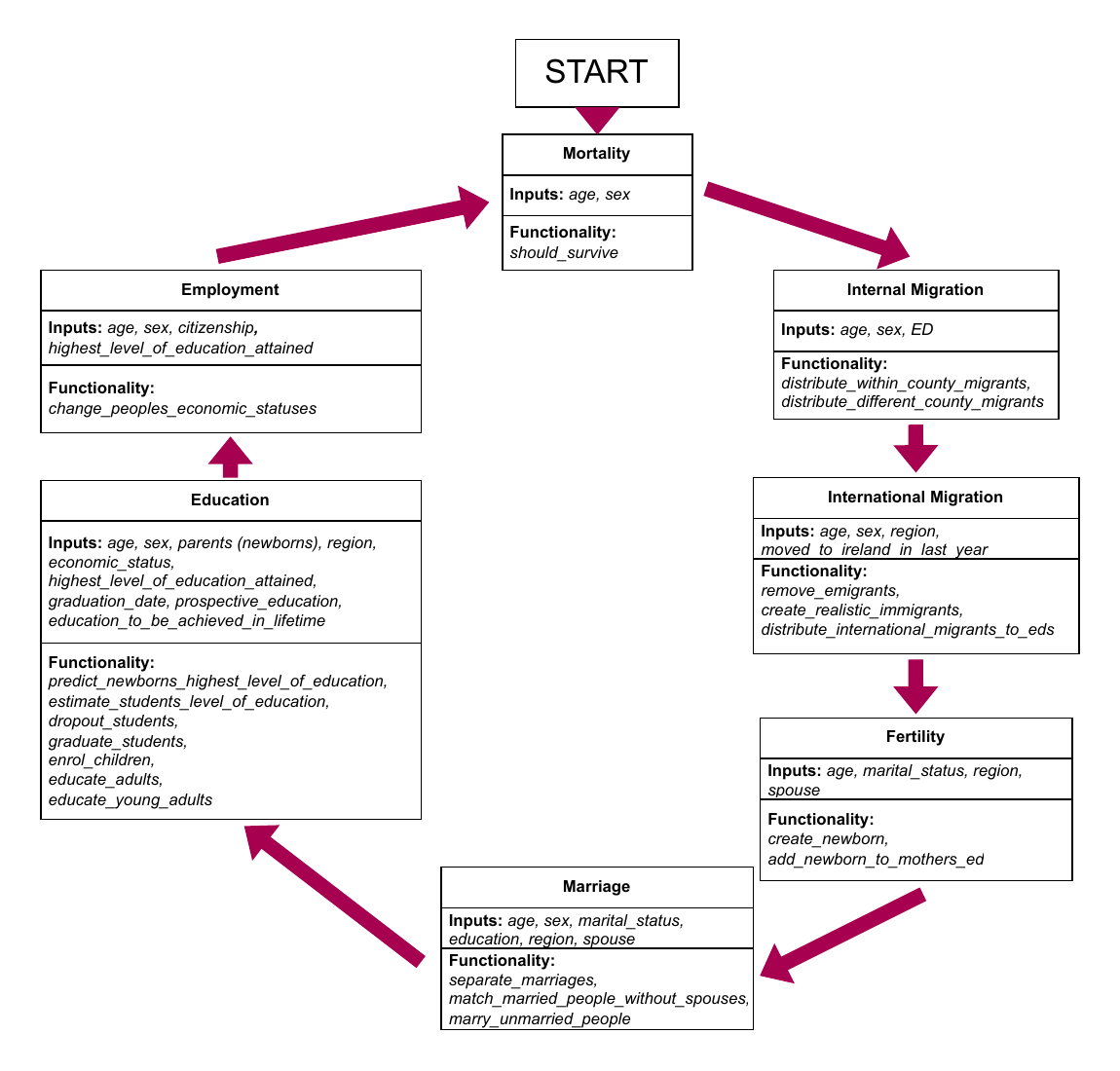}
    \caption{The order in which the annually-computed modules are calculated. All of the possible characteristics which could be used as inputs are outlined along with the major functions of each module.}
    \label{fig:moduleFlowchart}
\end{figure}

After the static population has been properly initialised, the microsimulation begins. The following sections outline, in order, the modules which are computed annually. The order of modules is chosen to allow for the most up-to-date information to be used for modules which rely on possibly new information from other modules. For example, the transition probabilities for a person’s economic status are dependent on their age, sex, citizenship, and education level. Therefore, the employment module is computed after every other module to allow for changes in age and education level in the same year. Figure \ref{fig:moduleFlowchart} illustrates this concept graphically and outlines the major functions included in each module. These functions are explained in more detail in the following sections.

\subsection{Mortality and Ageing}
\label{sec:mortality}
The mortality rates in 2016 for both sexes and each single year of age up to 105 years old are used to calculate each individual’s chance of dying in a year. 
In the CSO's regional population projections, life tables from 2017-2019 were estimated and used as the base mortality rates for 2022. These estimated tables were not made available, so they were estimated in SEMIPro by adjusting the rates until the initial numbers of deaths aligned closely to the values from the DCM.
There is then a linear decrease from 2.5\% improvement to 1.5\% improvement from 2022 to 2047. After 2047, the mortality rate improvement remains constant at 1.5\% per annum. However, no improvement is projected for those above the age of 100. Also, for individuals between 90 and 100, the annual improvement in mortality rate is scaled according to the person’s age’s distance between 90 and 100. For example, a 95 year old would experience a 1.25\% improvement in their mortality rate if the general improvement rate was 2.5\% as 95 is halfway between 90 and 100, and 1.25\% is halfway between a 2.5\% improvement and a 0\% improvement.

As in the CSO's regional projections, a regional variation is not applied to mortality assumptions or rates. This is due to ``the relatively small number of deaths in some regions, combined with the relatively small impact of regional differentials in the number of deaths'' \citep{cso2023projections}.

When a person dies, they are removed from their ED’s population and all references to them from other people are deleted. This means the deceased is removed from their children’s list of parents and their parents’ list of children. If the deceased has left a widow(er), the widow(er)’s ``spouse'' attribute is also deleted, and their marital status is set to Widow(er) (WID).

The ages of all of those who survive is then incremented by 1.

\subsection{Internal Migration}
\label{sec:internal_migration}

For all of the cases of immigration from within the same county, immigration from a different county and immigration from outside Ireland, a table of each ED’s population by usual residence 1 year prior is utilised. Each county’s total number of internal immigrants within each category is calculated which then allows us to calculate each ED’s proportion of its county’s immigrants from each category. Once the total number of emigrants for a county is calculated, the origin EDs of those migrants is sampled proportionally to each ED’s percentage of the county’s population. This may not be realistic, as it implies that people tend to emigrate from small and large EDs at the same rate. However, as there is no data for the number of emigrants for EDs, this approach is preferred over entirely random sampling of origins. Once an origin ED for a migrant has been sampled, the migrant is sampled from the origin ED's population. It is ensured that the sex and age group distribution of migrants is realistic by sampling proportionally to the observed distribution from 2022. These distributions are separate for inter- and intra- county migrants. Emigrants are then randomly assigned to a destination ED that has not exceeded its yearly number of internal immigrants.

\subsection{International Migration}
\label{sec:international_migration}
The number of international immigrants and emigrants in a year is controlled by the international migration scenario chosen at the beginning of the simulation. The CSO’s population projections explicitly state the assumed net migration for each scenario but the number of immigrants and emigrants is only implied. Therefore, there may be some minor discrepancies between the exact totals of migrants used by the CSO and estimated in this microsimulation. After the year 2032, the number of emigrants for the M1 and M2 scenarios is set to 50,000. The number of emigrants for the same period in the M3 scenario is set to 60,000. The number of immigrants after 2032 is set to 95,000, 85,000 and 70,000 for M1, M2, and M3, respectively. 

The first step in sampling international emigrants is to group the population by region. The number of emigrants from each region is calculated according to the previously observed regional proportions of emigrants. Emigrants from the region are then sampled proportionally to their sex and age group. These proportions are calculated by getting the average distribution of international emigrants over the period 2017-2022 (inclusive), as in the CSO's projections. Emigrants are removed from their origin ED’s population and references to them from other people are removed as in the case when people die. However, if the emigrant has a spouse, that spouse is given a spouse attribute of ``Outsider'', and their marital status remains as MAR.

As is the case for international emigrants, the age and sex of international immigrants is sampled according to the proportions of those characteristics observed in international immigrants to Ireland between 2017 and 2022 (inclusive).
Realistic distributions of socioeconomic characteristics for recent international immigrants is achieved using the 'Moved to Ireland in the last year' characteristic mentioned in Section \ref{sec:data}. The individuals with a value of ``True'' in this characteristic are saved at the start of the microsimulation. Then, international immigrants are assigned realistic characteristics by using a person of the same age group, sex, and citizenship from this population as a donor for each of the other characteristics. The donor's characteristics from the year they immigrated rather than their current characteristics are used. The destination ED of international immigrants is randomly sampled according to each ED’s proportion of the total number of people whose usual residence was outside of Ireland one year before the census.

\subsection{Fertility}
\label{sec:fertility}
In 2022, the TFR in Ireland was approximately 1.55. This study follows the CSO’s assumption that the overall TFR will decrease to 1.3 by 2038 and remain constant thereafter. Fertility rates are dependent on the mother’s age group, region of residence, and - if the mother is older than 24 - marital status. The reason for not including marital status as a predictor for those younger than 25 is that the low number of total births to women of those age groups was determined to be creating unreliable results (such as the lone married 19 year old in a region having 5 children in one year). The distributions for these rates are taken from the census data for 2022 and then decreased in each year following the general decrease in TFR. Once the number of births to mothers within a 5-year age group in a region has been calculated, mothers of the correct age are sampled randomly from the region’s population. In keeping with the current multiple pregnancy rates, each mother is given a $1.8\%$ probability of having twins. Triplets, quadruplets, etc., are not implemented in the microsimulation because of their extremely low prevalence. Once a mother has been assigned one or two newborns, she is removed from the sampling pool to disallow any further births to her in the year.

Newborns are assigned an age of 0 and a randomly chosen sex. The sex of the newborn is not weighted because Ireland has an approximately equal split between male and female births. The newborn’s marital status is set to SGL; their economic status is set to NA along with their education level. A citizenship value of ``IE'' is given to the newborn and their ``resided outside Ireland one year ago'' characteristic is set to False. They are assigned to the same ED’s population as their mother. This assumes that a newborn with parents living in separate EDs will move in with their mother. If the mother of the newborn is married, the mother’s spouse is assumed to also be a parent.

\subsection{Marriage}
\label{sec:marriage}
The only data tracked on divorces and separations in Ireland is from the courts system and gives the total number of divorces and separations granted annually. Therefore, a ``separation rate'' is calculated at the beginning of the simulation. This rate represents the number of separations and divorces proportional to the total number of married people. Then, in every year of the microsimulation, married couples are split randomly until approximately the correct number of separations according to this rate has been reached. Each individual in the couple’s marital status is updated to a value of ``Separated'' (SEP) and their spouse attribute is deleted.

Marriages are simulated using the same methodology as in the initialisation of the simulation except matching is done between individuals in the same region rather than those in the same ED. The marriage rate is assumed to be 4 marriages per 1,000 people, in keeping with the current rate. People to be married are sampled from all people without the marital status of MAR. This means that it is possible for those who are separated or widowed to re-marry. The distributions of the ages and sexes of those who were married in 2022 are used to achieve realistic candidates for marriage. The age and sex distributions of candidates for marriage are also unique across opposite-sex and same-sex marriages. Candidates are matched into couples using the same dissimilarity ranking based on age and education used in the initialisation of the population.

\subsection{Education}
\label{sec:education_module}

\subsubsection{Education to be achieved in lifetime}
\label{sec:education_to_be_achieved_in_lifetime}
The first step in the education module is to estimate the highest level of education that a newborn will achieve in their life. This is dependent on the newborn’s parents’ highest level of education achieved and the year of birth of the newborn. The parents' highest level of education is used to predict the broad level of education the newborn will achieve. The possible values for this broad level of education are ``Lower Secondary and Below'', ``Leaving Certificate/Post Leaving Certificate'' and ``Third Level''. The specific highest level of education to be attained in the newborn's lifetime is then sampled according to the proportion of the population in that year with each of the specific levels of education within the chosen broad level of education. For example, if the broad level of education chosen was ``Third Level'', the specific qualifications possible would be Undergraduate Degree, Postgraduate Degree and Doctorate. The probability of choosing e.g., Doctorate, would be the number of people in that year's population with a Doctorate divided by the total number of people with a ``Third Level'' qualification.  

\subsubsection{Dropping Out}
\label{sec:dropping_out}

Next, some students are dropped out from courses. Dropout rates at the Primary School level are set to 0 while those at the Secondary School level are set to 2.5\% initially and gradually decreased at the rate of improvement observed over the last 20 years. All dropout rates above the Secondary School level are set to the same rate as was observed in 2022. For courses with multiple year’s duration, the dropout rate is assumed to be uniform across each year of the course’s duration for ease of computation. Selection of which students should drop out is performed by prioritising those with a ``highest level of education to be attained in their lifetime'' equal to or less than the level of education they are currently studying for. 

When students are dropped out, their change of economic status is sampled according to the non-progression outcomes observed in 2022. These outcomes are determined by the level of education the student was attempting to achieve. A dropout whose next step is selected to be re-enrolment in education is re-enrolled in the first year of a course decided by their current level of education. If the dropout has a ``highest level of education to be attained in their lifetime'' attribute and this level has been surpassed, the option to re-enrol in education is eliminated as a possibility. 

\subsubsection{Graduation}
\label{sec:graduation}

Then, all remaining students’ graduation dates are checked and those with a graduation date of the current year of the simulation are graduated. Their highest level of education achieved is updated to the prospective level of education they were studying for. Graduates of Primary School and Lower Secondary are automatically progressed to the next level of the school system. The previously mentioned dropout rates for those levels of education are determined by the number of young adults with a lower secondary education or below, so students graduating from those levels and not continuing their education would skew the education rates incorrectly. As with students who have dropped out, each graduate’s next step is decided based on the level of education they were studying for. Continuing in education is removed as an option for those with a ``highest level of education to be attained in their lifetime'' at or below their current level of education. One point to note here is that the data on graduation outcomes is only available for ``young'' graduates but is applied for all graduates here. 
The rate of employment after graduation for adult learners is likely higher than those of younger graduates considering a large number of adult learners are studying to do their job better or improve their career prospects \citep{cso2024adulteducationsurvey}.
As data separating graduation outcomes by age is not currently available, we consider this as an improvement for the future.

\subsubsection{Enrolment}
\label{sec:enrolment}
Individuals with an age equal to the school enrolment age, assumed in this microsimulation to be 4 years old, are then added to the Primary School system. 

\subsubsection{Adult Learners}
\label{sec:adult_learners}
The final two steps in the annual education module calculations are to add those above the age of 17 back into education. The proportion of adults between the ages of 25 and 69 that were in formal education in 2022 is used to calculate the number of older students to be added in each region. In 2022, approximately 61\% of people between the ages of 18 and 24 had the economic status of student. This proportion is used to calculate the number of individuals in this age group to add back to the system. Individuals to be added to the education system are sampled based on the age and sex distributions of adult learners observed in 2022. Those who have not attained their ``highest level of education to be achieved in their lifetime'' yet are prioritised for re-entry into education. As in the case of young graduates continuing on in education, an adult learner's current highest level of education attained is used to stochastically determine their new prospective course.

\subsection{Employment}
\label{sec:employment}
The economic statuses of all individuals older than 15 who are not students (students are dealt with in the education module) are transitioned based on the individual’s age, sex, citizenship and education. The transition probabilities are determined based on the proportion of the 2022 population fitting in the same age, sex, citizenship, and education bracket with each economic status. 
Note that there is an absence of data in Ireland on current economic status based on the economic status in the previous year. Therefore, an employed person is equally as likely to become unemployed as an unemployed person of the same age, sex and education is to stay unemployed.

\section{Results}
\label{sec:results}

This section outlines the national, regional and socio-economic results for the microsimulation. In the interests of brevity, the M1 international migration scenario is used as a reference, except where otherwise stated. The corresponding results for the M2 and M3 scenarios can be found in the attached GitHub repository. The presented results for each scenario are calculated from the average results over 5 runs of the microsimulation.

\subsection{National Validation}
\label{sec:national_validation}


\begin{figure}[!htb]
    \centering
    \includegraphics[width=\linewidth]{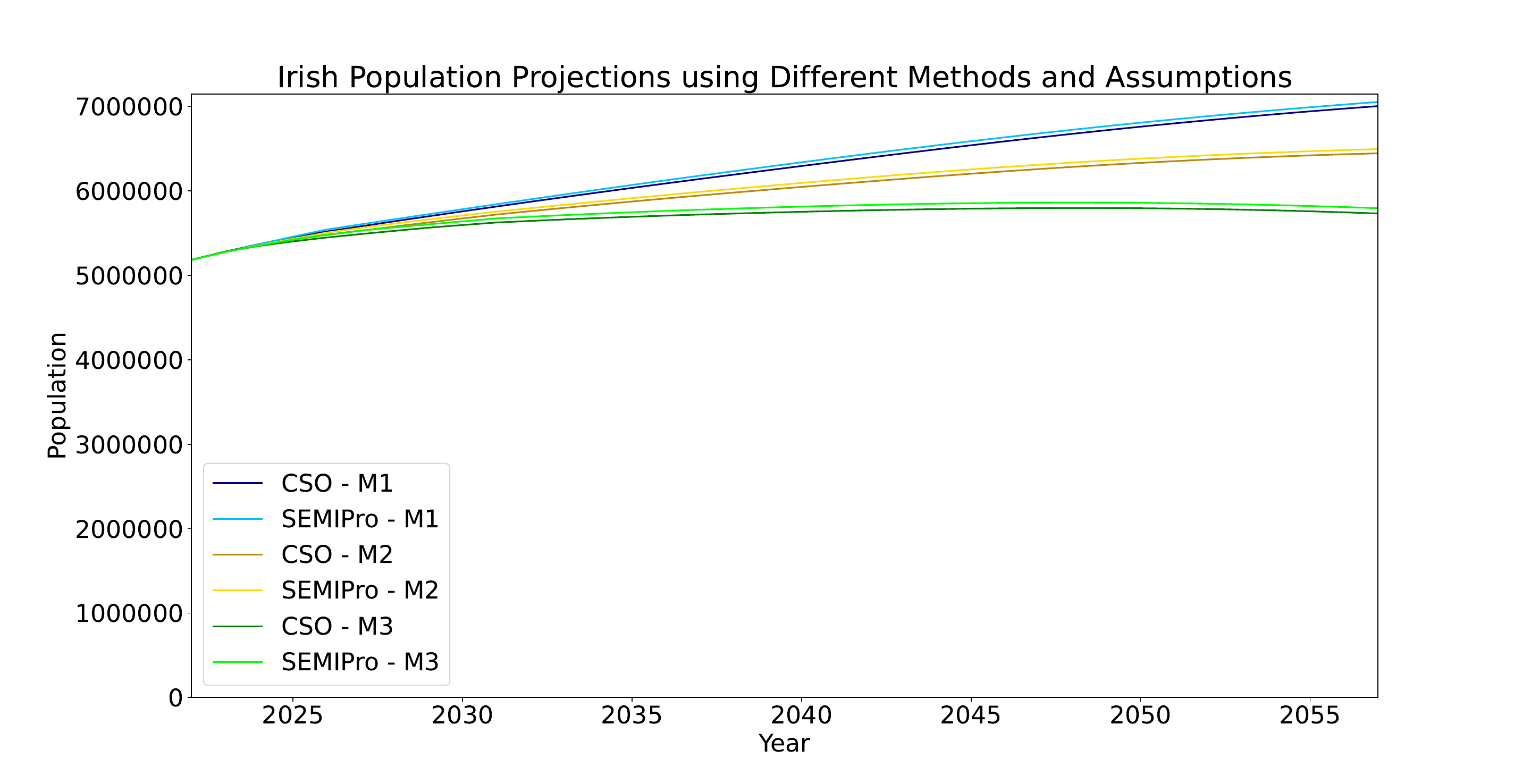}
    \caption{The changes in the population size for all three international migration scenarios using both microsimulation and DCM methods}
    \label{fig:annualPopulationsOursVsCSOs}
\end{figure}

This microsimulation approach is validated by comparing the size and composition of the final populations to those generated in the CSO's DCM population projections. As evidenced by Figure \ref{fig:annualPopulationsOursVsCSOs}, the populations generated by the microsimulation track the CSO's population sizes reasonably well.

\input{m1_population_comparison_table}

Table \ref{tab:comparisonTableM1} details the numerical differences in the structure of the populations between the two approaches for the M1 international migration scenario. The corresponding tables for the M2 and M3 scenarios are available in Appendix \ref{sec:dependency_ratios}. Here, the young dependency ratio describes the number of people aged 0-14 divided by the number of people aged 15-64 (approximately the ``working'' age range) expressed as a percentage. The old dependency ratio describes the number of people aged over 65 divided by the number of people in the same ``working'' age range. 
The minor differences between results in Table \ref{tab:comparisonTableM1} can be explained by a number of factors. Firstly, the CSO present their international migration assumptions as a table of the average number of annual immigrants and emigrants over 5-years periods from 2022 to 2057. The exact numbers for each individual year could easily differ between their simulation and this paper's. International migration is also also by far the largest driver of population change in these simulations, so small differences could have relatively large effects. Similarly, the decline in the TFR from 1.55 to 1.3 from 2022 to 2038 could be implemented as a linear, exponential or stepped decay which again would lead to differences in the number of births in each year.  
Considering the above, it was decided that the populations resulting from these microsimulations matched the CSO's populations closely enough to begin deeper analysis of the rest of the projection results.

\subsection{Regional Validation}
\label{sec:regional_validation}

\begin{figure}
    \centering
    \includegraphics[width=\linewidth]{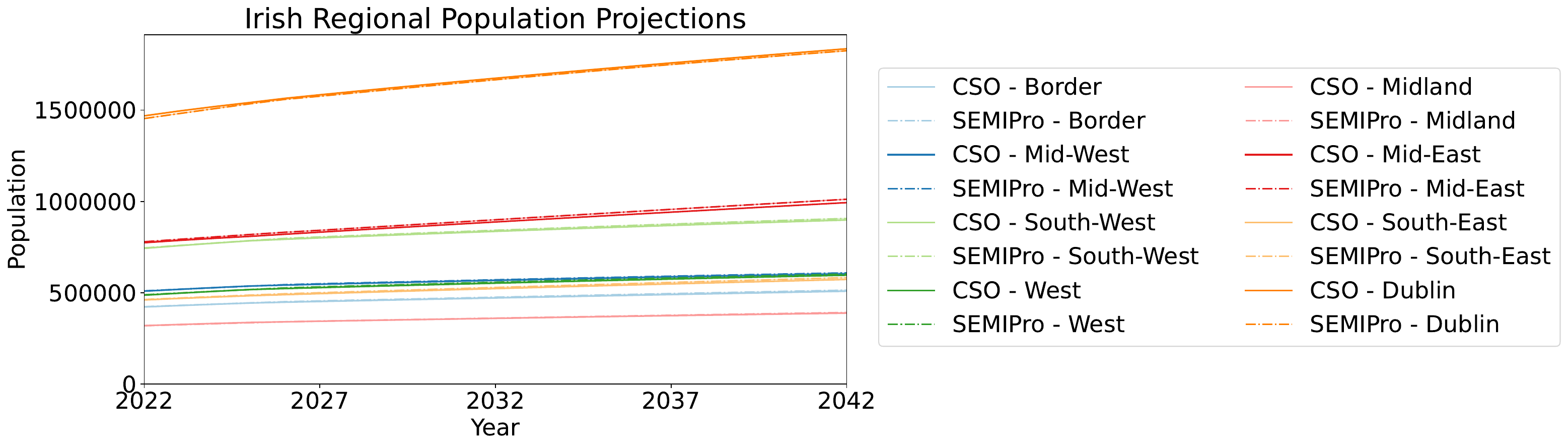}
    \caption{Changes in population size at the regional level}
    \label{fig:regionalPopulationChanges}
\end{figure}

The CSO's Regional Population Projections allow us to compare the populations generated by SEMIPro to the DCM method from 2022 to 2042. The changes in population in the M1 scenario for each region for both methods are given in Figure \ref{fig:regionalPopulationChanges}. As in the national case, there is close agreement between the SEMIPro and DCM populations. One thing to note in these plots is that the range of populations for each region over each of the 5 microsimulations is plotted, but is almost imperceptible. This emphasises the reproducibility of the microsimulation approach at the regional level.

\begin{figure}[!htb]
    \centering
    \includegraphics[width=\linewidth]{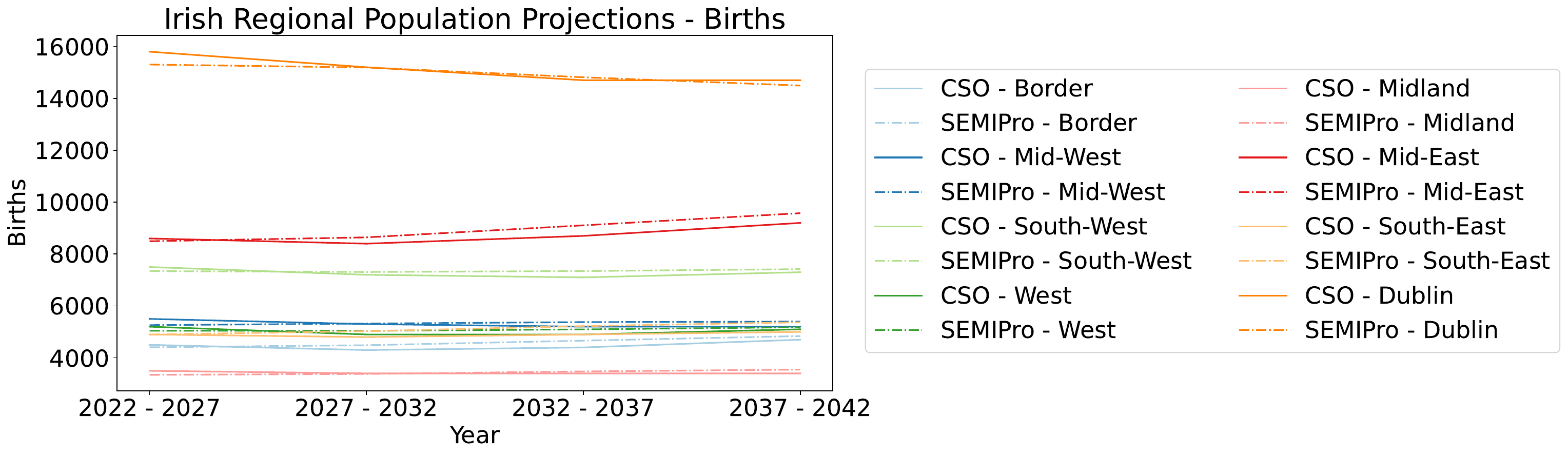}
    \caption{A comparison of the average births per intercensal period in each region using both SEMIPro and DCM}
    \label{fig:regionalBirths}
\end{figure}

\begin{figure}[!htb]
    \centering
    \includegraphics[width=\linewidth]{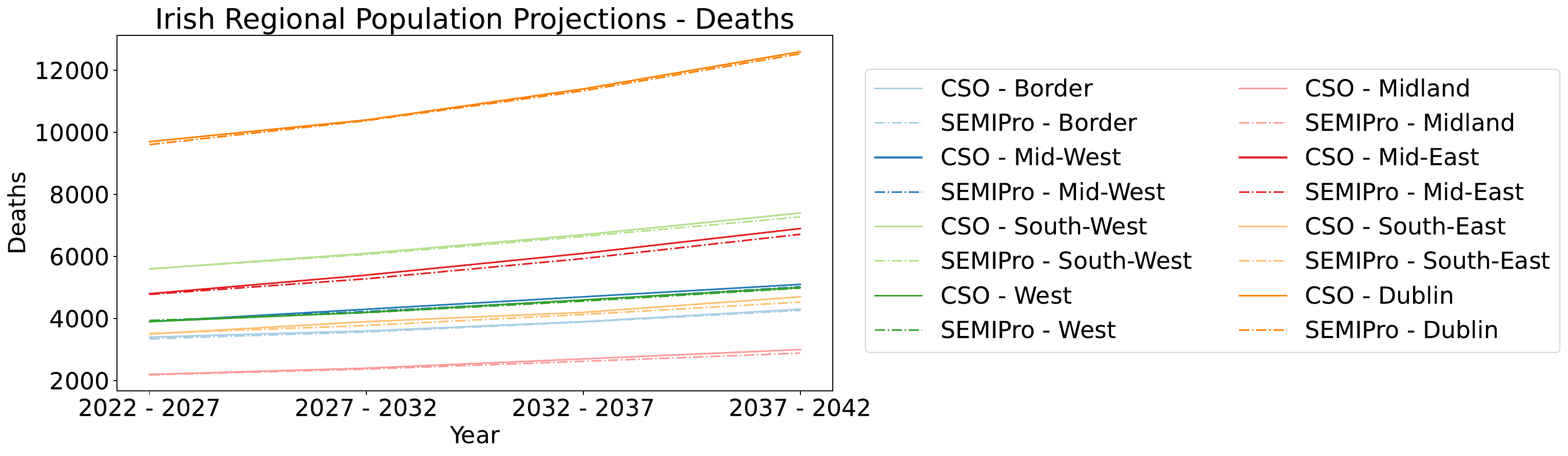}
    \caption{A comparison of the average deaths per intercensal period in each region using both SEMIPro and DCM}
    \label{fig:regionalDeaths}
\end{figure}

\begin{figure}[!htb]
    \centering
    \includegraphics[width=\linewidth]{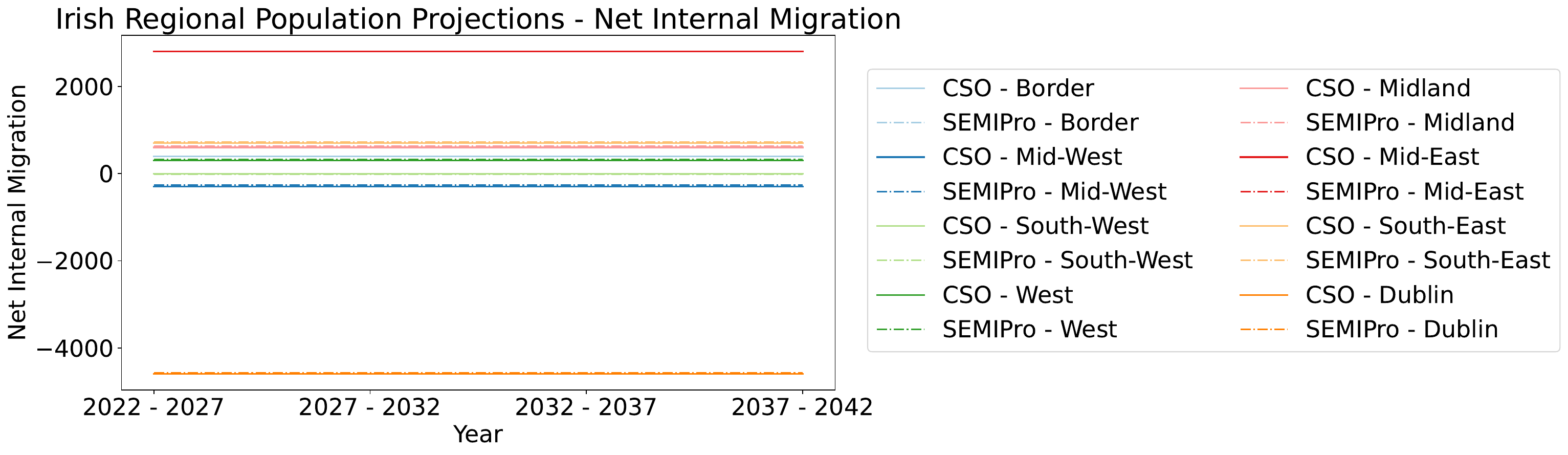}
    \caption{A comparison of the average net internal migration per intercensal period in each region using both SEMIPro and DCM}
    \label{fig:regionalNetInternalMigration}
\end{figure}

\begin{figure}[!htb]
    \centering
    \includegraphics[width=\linewidth]{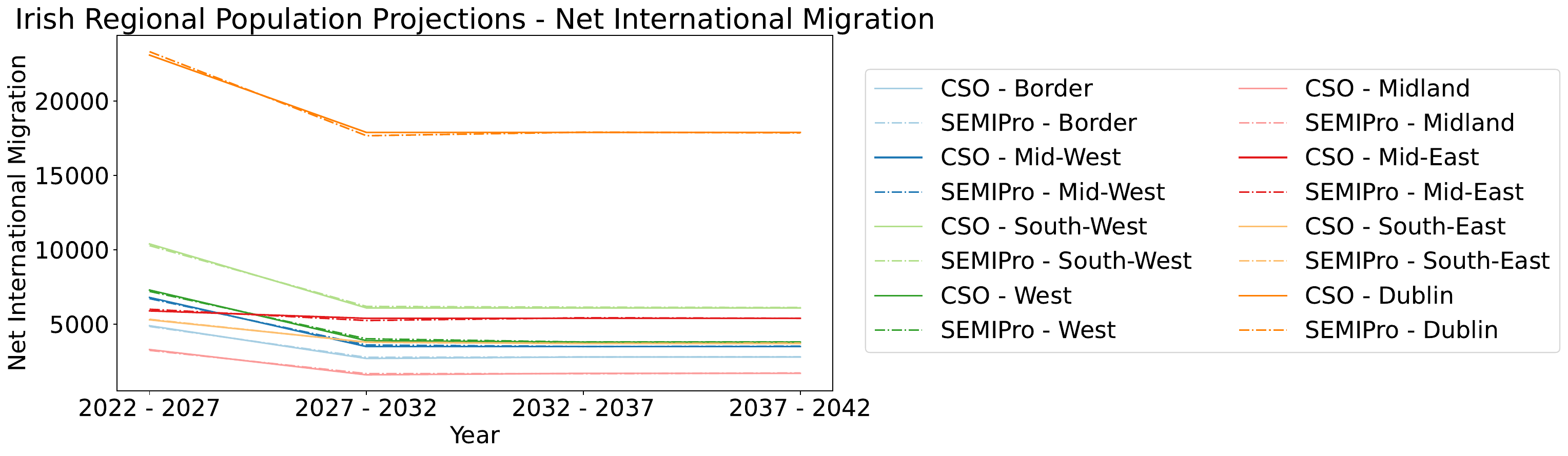}
    \caption{A comparison of the average net international migration per intercensal period in each region using both SEMIPro and DCM}
    \label{fig:regionalNetInternationalMigration}
\end{figure}

In order to understand why there are differences between the SEMIPro and DCM populations, even if they are small, the number of births and deaths as well as the net internal and international migration for each region are compared to their DCM counterparts. These comparisons are illustrated in Figures \ref{fig:regionalBirths}-\ref{fig:regionalNetInternationalMigration}.

In both types of migration, it is evident that there is essentially a one-to-one match between the SEMIPro and DCM averages. The reason for this is that these statistics are directly controlled by the assumptions of their relevant methods. For example, although the exact number of international immigrants to Ireland each year is not detailed explicitly in the CSO's projections, a realistic approximation of this value can be estimated. Then, the national net international migration will be known from the assumptions (as described in Section \ref{sec:scenarios}), from which it is trivial to calculate the required number of international emigrants. These national figures can then simply be proportionally spread across each of the regions, ensuring a nearly exact match with the DCM results. The reason for any small deviations in these migration figures is due to alignment to the CSO's rounded figures. For example, due to rounding, the sum of the CSO's net international migrations for each region is sometimes slightly off from the figure stated in the assumptions e.g., +45,000 in 2032-2037, because each region's figure is rounded to the nearest thousand.

There are slightly larger differences evident in the birth and death statistics which are controlled by rates, rather than number of occurrences. For example, the number of deaths per year in SEMIPro remains just slightly lower than the number from DCM. It was decided not to alter the mortality improvement implementation for a number of reasons including the relatively small differences in deaths and the fact that the trend in deaths is matched closely. Small deviations are present in the first intercensal period, which may indicate a difference in the characteristics of the initial population between the two methods. The CSO state that initial mortality rates were calculated by generating estimated life tables for Ireland from 2017-2019 data. Unfortunately, these life tables and this data were not available here so estimates were calculated by running SEMIPro with different starting mortality rates and searching for a close match to the correct number of deaths.

As mentioned in Section \ref{sec:national_validation}, the CSO's fertility assumption can be implemented many ways. Therefore, the small differences in the regional numbers of births are to be expected. One potential reason for these variances could be a difference in the way in which fertility rates were diminished. In this first version of SEMIPro, the same decline in fertility rate is implemented for each age group of mother. However, the CSO note that their assumption of the TFR decreasing from 1.55 to 1.3 is based on a number of factors including the increasing average age of first-time mothers. Therefore, some sort of scaling to decrease young mothers' fertility rates more than older mothers' may have been implemented in the DCM without being stated outright.

Although the above paragraphs may have zeroed in on potential reasons for the minor differences in the number of births, deaths and the net internal and international migration between SEMIPro and the DCM, it should be emphasised that the SEMIPro results are extremely encouraging. Firstly, the differences that do exist are minor, especially when viewed relative to the total number of occurrences of each of the statistics. Secondly, and likely more importantly, the regional trends in each statistic are almost entirely replicated. This suggests that if a user would like to propose their own different starting conditions, e.g., perhaps ``healthier'' initial mortality rates, they can have confidence in the resulting trends in numbers of deaths.

\subsection{Uncertainty of Results}
\label{sec:uncertainty}

Uncertainty of results is a problem which is widespread across all microsimulation. In an attempt to quantify this uncertainty, the approach outlined by \citet{jia2023norway} of running the simulation multiple times for a scenario and calculating the average and dispersion of the results is adopted here. The average of 5 runs are presented here.


\begin{figure}
    \centering
    \begin{subfigure}{0.4\textwidth}
    \includegraphics[width=\textwidth]{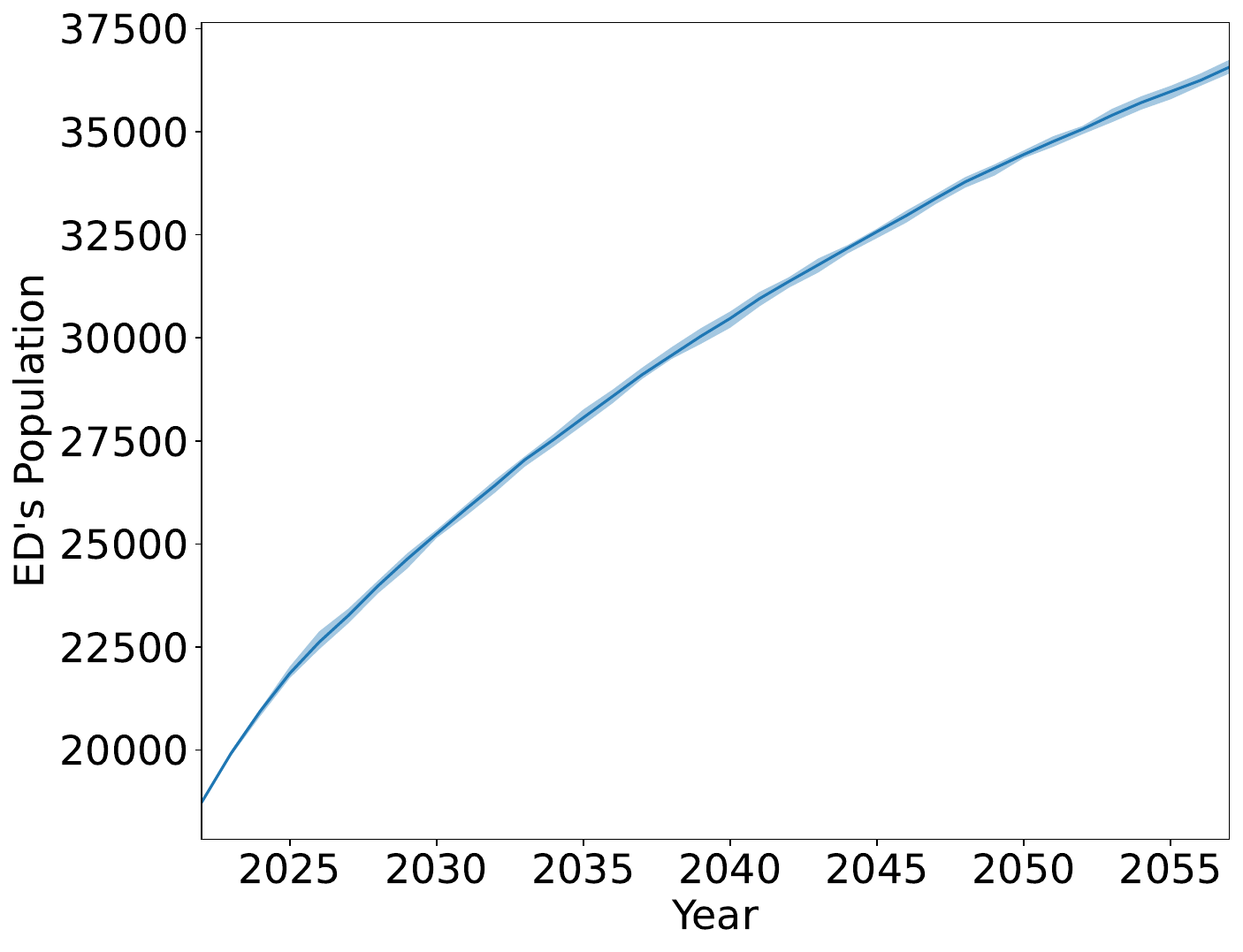}
    \caption{Largest ED}
    \label{fig:largest_ed}
    \end{subfigure}
    \hfill
    \begin{subfigure}{0.4\textwidth}
    \includegraphics[width=\textwidth]{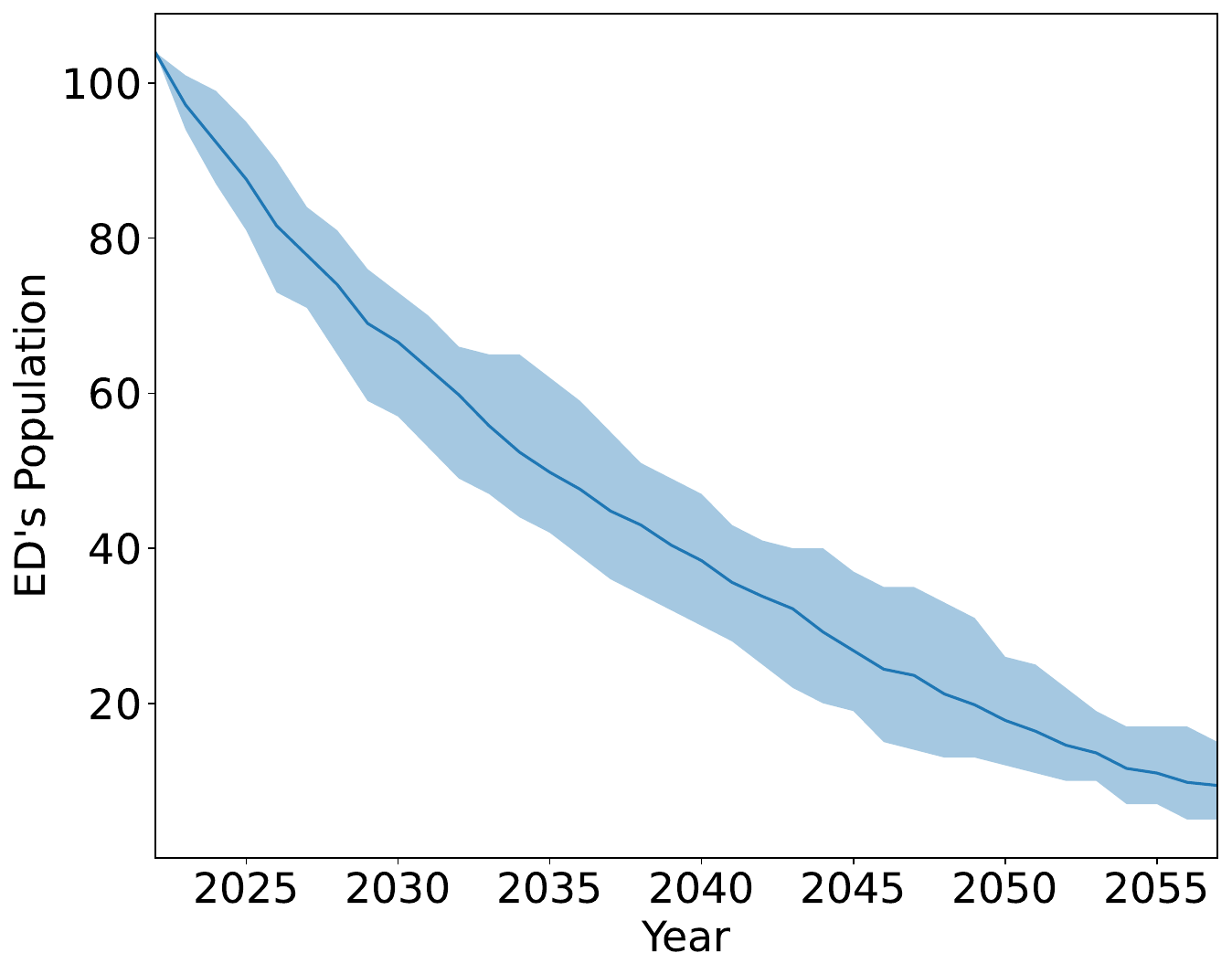}
    \caption{Smallest ED}
    \label{fig:smallest_ed}
    \end{subfigure}
    \hfill
    \begin{subfigure}{0.4\textwidth}
    \includegraphics[width=\textwidth]{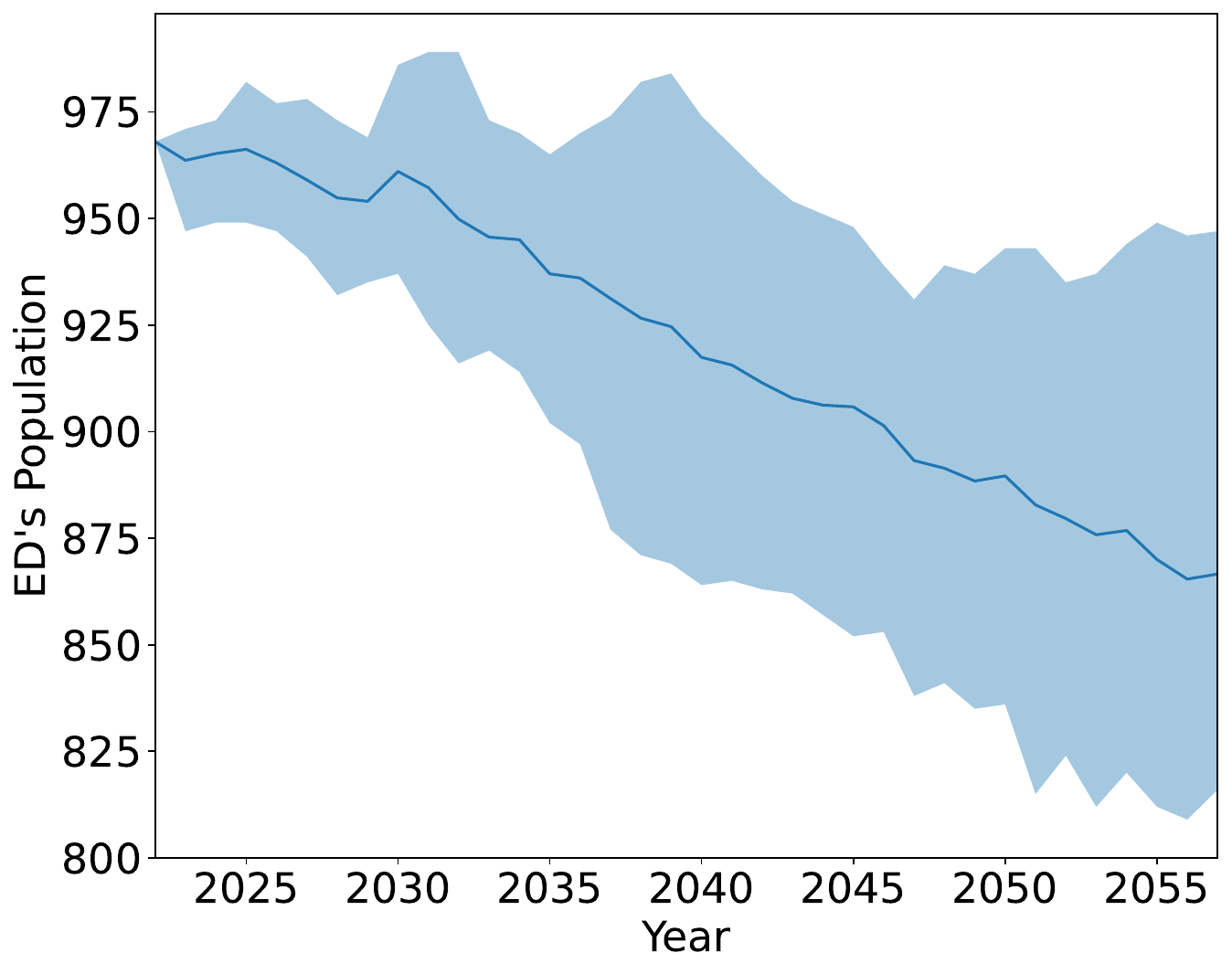}
    \caption{Median ED}
    \label{fig:median_ed}
    \end{subfigure}
    \caption{Changes in populations for the largest, smallest and median ED. For each year, the mean population is plotted on the solid blue line, while the shaded areas surrounding the line represent the year's range of simulated populations.}
    \label{fig:stability_of_eds_pops}
\end{figure}

In Figure \ref{fig:stability_of_eds_pops}, the changes in populations from 2022 to 2057 for 3 reference EDs are plotted. The largest, smallest, and median ED were chosen according to the mean 2057 population for each ED. Although Bayesian methods were not implemented in this study, the results display similar trends to those using Bayesian methods \citep{jia2023norway}, insofar as absolute deviations being directly proportional to ED population size, and relative deviations being inversely proportional to ED population size. The smallest ED's range of populations is $[5, 15]$ with a mean of 9.4 while the largest ED's range of populations is $[36413, 36745]$ with a mean of 36567. These results are comparable to Bayesian methods, where the 90th percentile prediction interval for the smallest municipality in Norway is $[4, 15]$ and the average is approximately 9 \citep{jia2023norway}. Of course, a large portion of the stability of these results can be described by the deterministic nature of the number of international immigrants to each ED each year. A discussion of incorporation of stochastic modelling into SEMIPro is given in Section \ref{sec:limitations}.

Outside of the numbers of international immigrants to an ED, essentially everything else which may affect an ED's population size is determined stochastically. For example, on the other side of international migration, international emigrants are sampled from a region's entire pool of people in a cohort. Births are also sampled from a regional population. Inter- and intra- county migrants are sampled from counties' entire populations. Therefore, the relative stability of the results above is encouraging and suggests that extension of the model with features like Bayesian models for international migration will be relatively seamless.

\subsection{Overall Results}
\label{sec:overall_results}

\subsubsection{ED Sizes}
\label{sec:ed_sizes}
The CSO's analysis of the age and sex composition of the population in 2057 is comprehensive so this paper will not labour the point about the same topics. Instead, the focus in the following sections will be on results for individual EDs, results for the new socioeconomic characteristics and a combination of the both of these. In the interest of concision, the results presented in all of the following sections will be for the ``M1'' International Migration except where otherwise stated.


\begin{figure}[!htb]
    \centering
    \includegraphics[width=\linewidth]{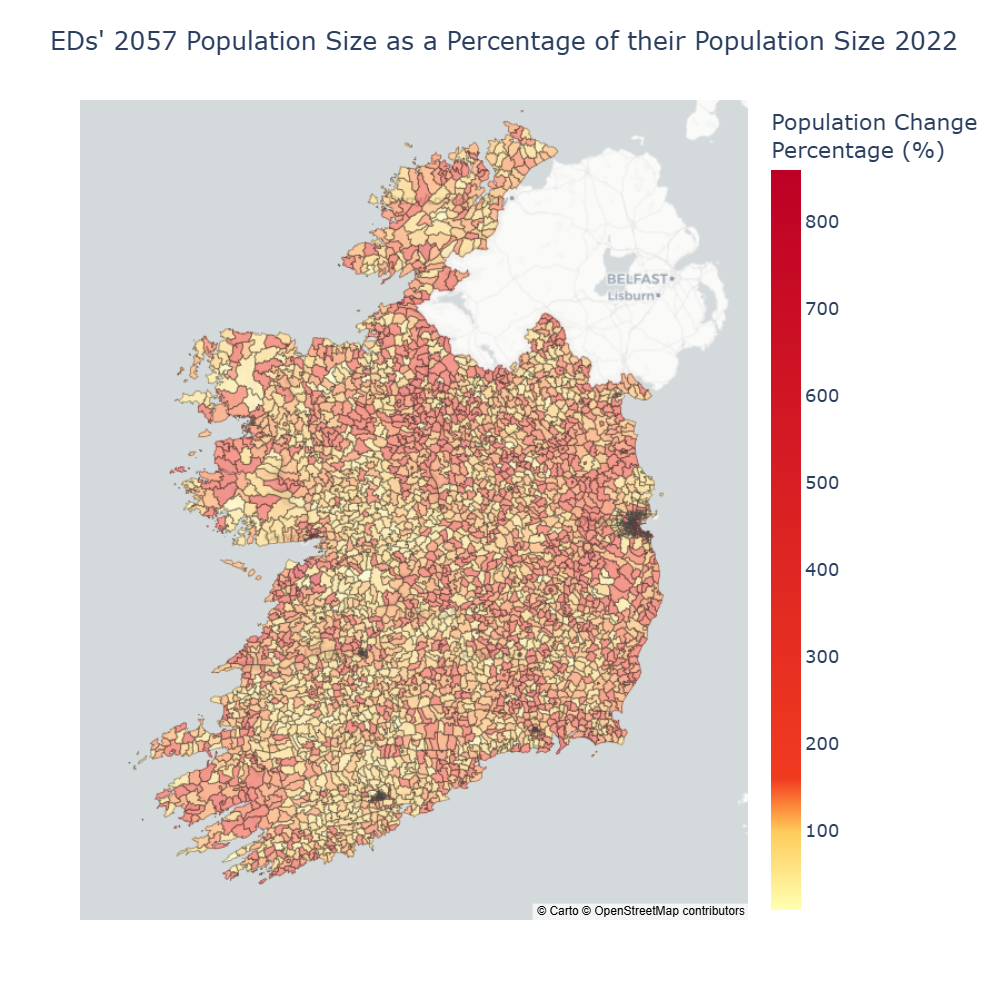}
    \caption{The size of EDs in 2057 relative to their size in 2022}
    \label{fig:percentageChangeChoropleth}
\end{figure}

Figure \ref{fig:annualPopulationsOursVsCSOs} shows the relative changes in ED sizes over the 35 years of the simulation. An ED with a value of 300\% in the figure means its population will have tripled in size in the simulation. Clearly, there is a wide range of changes in the relative population sizes, from a minimum of approximately $9\%$ to a maximum of approximately $861\%$. The ED at this minimum relative size is a rural area called Derrylaur in Co. Galway. 
It is projected to decline from a population of of 99 to an average population of just 9 people. While this almost complete decline of this ED's population may not be likely in reality, the simulation logs allow us to track why the results came about. 
In 2022, the average age of the female population in Derrylaur was approximately 48 years old, quite a bit older than the national average of approximately 38 years old. 
This relative lack of potential mothers leads to there only being an average of approximately 3 births in the ED over the entire simulation period. Furthermore, in the year prior to the 2022 Census, Derrylaur received no internal or international immigrants. As mentioned in Sections \ref{sec:internal_migration} and \ref{sec:international_migration}, emigrants out of EDs are sampled according to the ED's share of its county's population. 
This leaves a small but always present probability of emigrants leaving Derrylaur with no chance of any immigrants to replace them. Again, the scenario of an ED with enough housing for almost 100 people in 2022 declining to only approximately 9 people in 2057 is probably not realistic. However, the usefulness of this microsimulation is that policy makers and other researchers can analyse which EDs are at risk of serious decline, and potentially even test mitigation strategies in the simulation themselves.

The largest ED in 2057 is Maynooth in Co. Kildare, with an average population of 36,654. Interestingly, Maynooth starts the simulation as only the 15th largest ED in the country, with a population of 18,238. Examination of the factors influencing Maynooth's growth reveals almost an exactly opposite picture to that of Derrylaur. The mean starting age is quite young at approximately 33 years old and the ED takes in a large proportion of internal migrants both from within Kildare and from other counties. Kildare's proximity to Dublin means that it receives by far the largest number of migrants from the capital in this net Dublin outflow internal migration scenario. Finally, Maynooth being situated in the Mid-East region means it has a much higher TFR than the national average. The consequences of Maynooth, a university town, experiencing such significant growth could have significant implications for its housing market, its transport connections to Dublin, and many other factors which the government and local councils will have to account for.

One point to re-emphasise here is that all of the above results are occurring under ``base case'' assumptions with the majority of the data informing the microsimulation being copied from 2022 values. Even disregarding the type of simulation uncertainty outlined in Section \ref{sec:uncertainty}, there is a large degree of uncertainty to whether these assumptions will hold true or not. Therefore, it is recommended that SEMIPro users monitor and update assumptions as new data becomes available.

\subsubsection{Immigration During the Simulation}
\label{sec:immigration_results}
As previously mentioned, international migration is projected to be the largest driver of population change in Ireland in the period of 2022-2057. The percentage of the population in 2057 who have moved to Ireland in the period 2022 to 2057 is projected to be approximately 19.86\%. 
Note that approximately 32\% of international immigrants to Ireland between 2017 and 2022 held Irish citizenship, so this number will include Irish citizens returning home.

\begin{figure}[!htb]
    \centering
    \includegraphics[width=\linewidth]{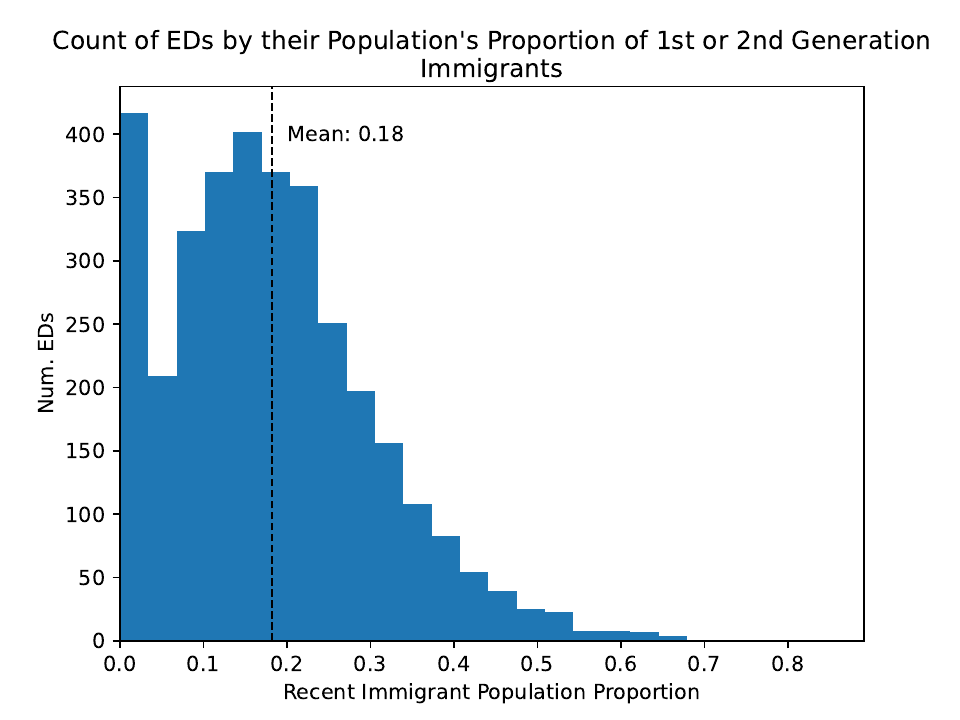}
    \caption{The number of EDs with different proportions of recent immigrants in their populations}
    \label{fig:immigrantHistogram}
\end{figure}

Figure \ref{fig:immigrantHistogram} displays the distribution of proportions of EDs' populations that are either recent immigrants to Ireland or the child of such a recent immigrant. The overall distribution shape is unlikely to be surprising, with the majority of EDs situated around the mean and a gradual tail off at higher proportions. There are two interesting points, however. The spike at 0 highlights the relatively large number of EDs who did not have any international immigrants in 2021, and thus are projected not to have any during the course of the microsimulation either. On the other end, there is a small spike above the proportion of 0.8. The ED projected to have the highest proportion of recent immigrants and their parents is Killygarvan in Co. Donegal, where that proportion is nearly 85\%. Although Killygarvan begins with a population of just 120, it has an average net international migration inflow of $\approx 9.5$ per year. This high relative inflow of international immigrants is contrasted with the ED's average net internal migration which is an outflow of $\approx5.5$ people per year. 

The reason for presenting results for recent immigrants is that, as outlined by \citet{mcginnity2025esriIntegration}, there are differences in socioeconomic outcomes for migrants and non-migrants. In particular, the rate of employment as well as the rate of labour market participation for migrants are higher than the Irish-born population. Also, the tertiary education rates for working age migrants are higher than their Irish counterparts. Finally, both first- and second- generation migrants in Ireland score lower on the Programme for International Student Assessment subjects than their non-migrant counterparts (which is largely due to socioeconomic differences and spoken language at home)\citep{donohue2023irishStudentPerformance}. Again, the results presented by \citet{mcginnity2025esriIntegration} related specifically to migrants with a place of birth other than Ireland, but considering the relative lack of data, we consider the SEMIPro results a useful point of reference for local planners.

\subsection{Socio-economic Results}
\label{sec:socioeconomic_results}

\subsubsection{Education}
\label{sec:education_results}


\begin{figure}[!htb]
    \centering
    \includegraphics[width=\linewidth]{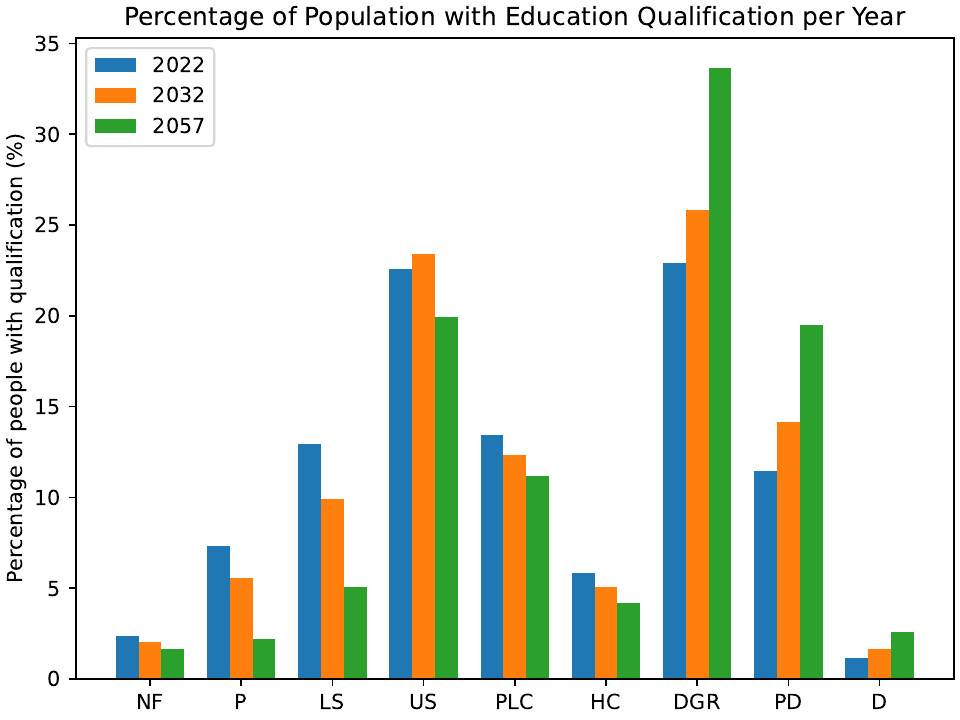}
    \caption{The proportions of the adult population with various levels of education for 2022, 2032 and 2057}
    \label{fig:educationComparison}
\end{figure}

Ireland has seen a significant increase in the levels of education of its population in recent years \citep{cso2024educationprogress}. Figure \ref{fig:educationComparison} emphasises that this trend is projected to continue and even accelerate more quickly. There are decreases in the percentage of adults with all levels of education below third level. Meanwhile, the three levels of university education are projected to increase correspondingly. The National Strategy for Higher Education to 2030 \citep{deptEducation2011heStrategy} highlighted capacity as one of the main challenges that the Irish education system was set to face and this significant demand for third-level courses will surely prove to continue to challenge that capacity.  

\subsubsection{Economic Status}
\label{sec:economic_status}

The economic status module in the microsimulation is based on age, sex and education level as mentioned in Section \ref{sec:employment}. The previous section outlined how Ireland's overall level of education is set to increase significantly over the simulation period. Therefore, we would expect the employment levels to reflect that improvement in education. The unemployment rate in 2022 was approximately 6\%. This rate is projected to decrease relatively to approximately 4.7\% in 2057. Interestingly, the unemployment rate is projected to be approximately 4.2\% in 2032, lower than the 2057 value. One theory for this behaviour is that in 2032, Ireland will have seen high net migration for the previous 10 years. As previously mentioned, recent immigrants to Ireland have higher rates of employment than the resident population \citep{mcginnity2025esriIntegration}. The decreasing unemployment from 2022 to 2032 may then be explained by the high inflow of people into Ireland in they period. On the other hand, as this net international migration begins to decline, the effects on the unemployment rate decline too. Again, this improvement in employment is reliant on there being enough jobs for the population which is dependent on many external factors to this simulation.

\section{Limitations and Future Work}
\label{sec:limitations}
As is always the case with projections of any kind, the relevance of the above results partly depends on the accuracy of the underlying assumptions of the simulation. The addition of the education, marriage and employment modules brings with it a plethora of assumptions, some of which may be turn out to be accurate and some which may not. The aim of attaching the simulation source code and explicitly outlining all of the added assumptions (Appendix \ref{sec:assumptions}) is that a researcher who may believe an alternate scenario is likely can easily alter the code to generate results for their set of assumptions.

At various points throughout this paper, an absence of data has been pointed to as a reason for non-inclusion of a feature. With regards to fertility, there are tables available detailing the number of births to women disaggregated by previous number of liveborn children. These tables could be used to more accurately sample realistic mothers provided that the previous number of children was available in the microsimulation. When a birth occurs in the simulation, it is trivial to append the child to the mother's previous list of children, so this characteristic would be as simple as checking the length of the list of children. However, the issue lies in the static population procedure used to generate the initial population. Previous number of liveborn children is not tracked in the the LFS or SAPS. In the LFS, the number of children linked to a family unit could be used as a proxy for the desired characteristics but this is not an exact equivalence. Likewise, in the SAPS, the number of families by counts of children represents a similar value. However, neither of these values take into account children not living with the mother.

Another module hampered by a lack of data is internal migration, where data on migration flows are not disaggregated by economic status. The ability to separate migration by economic status would be particularly useful for EDs in Ireland's college towns/cities such as Galway, Maynooth, Cork, etc. In the same vein, even if age specific county-to-county migration counts were available, these may be able to implicitly move students to college towns, given that students are typically young, and college towns are major hubs for internal immigration, generally. The economic status module is another feature which would benefit from disaggregation by just one more characteristic, namely previous economic status. Unfortunately, the LFS does not track individuals across quarters (with an ID or some equivalent), otherwise individuals' economic statuses in the first and fifth quarter of the survey could be used to inform this decision making. Finally, owing to the relatively low number of Irish citizenships granted as well as the complexity of the model already, the granting of Irish citizenships was not implemented here. A module implementing a possible change of citizenship adhering to the Irish eligibility rules may be a desirable addition in the future.

As mentioned in Section \ref{sec:intro}, Bayesian sampling of small area parameters is a prominent research area in small area research. Therefore, implementation of Bayesian estimators of statistics such as international immigration flows to specific EDs, internal migration flows into and out of EDs, and economic status changes by age, sex, education and citizenship would enhance the model and is planned for future work. Implementation of Bayesian estimators in this first iteration of the model was considered, but due to the size and complexity of the model as is, it was decided to postpone measures of stability and Bayesian modelling until the second iteration of the model, after receiving feedback and advice regarding this iteration. This second iteration of the model is planned for the near future. In terms of where small populations are present in SEMIPro decisions, migration and economic status changes are the main areas, as previously mentioned. Mortality is disaggregated by age and sex at the national level. Fertility is disaggregated by age and region. Marriages are created amongst the national population. Both of these modules will therefore have reasonably sized populations. There may be small populations present in the education module when deciding on the outcomes for students dropping out, especially from both low and very high levels of education. The same may be true for students graduating from doctorates. Finally, small populations may also be present when dealing with older adults returning to education.

Another factor related to scarcity of data, validation and Bayesian modelling is that statistics and even a DCM at more granular levels than regional statistics would be highly beneficial to this microsimulation model. In Ireland, regions are composed of counties which themselves are composed of EDs. However, as evident in this paper, data is often only available at the regional level. The same is true of DCM results. An example of a module which would benefit massively from more granular data is fertility. If fertility rates for each ED were available, much more geographical heterogeneity could be captured. Furthermore, Bayesian modelling could be employed as in Norway \citep{leknes2021norwaySmallAreas} to treat the fertility rates of small EDs with more scepticism than those of larger EDs. The same sentiment also applies to other modules in SEMIPro where granular geographical data is also not available.

\section{Conclusion}
\label{sec:conclusion}

This paper has introduced a dynamic microsimulation model developed for Ireland, SEMIPro, capturing individual-level demographic and socioeconomic transitions over a 35-year period from 2022 to 2057. By simulating births, deaths, internal migration, and international migration, and by tracking key characteristics such as age, sex, marital status, citizenship, whether the person resided in Ireland one year ago, highest level of education achieved, and economic status, the model offers a detailed and flexible representation of population dynamics.

The model enables the examination of long-term demographic trends and the distributional impacts of various policy scenarios. Its ability to represent individual heterogeneity and simulate life-course events makes it a valuable tool for both researchers and policymakers seeking to understand the complex interactions between demographic behaviour and socioeconomic conditions.

One of the key results from the microsimulation is Maynooth's growth from the 15th largest to the largest ED in terms of population in the country. Assuming that the average number of people per private household remains at the 2022 national level of 2.74 and that the ED's occupancy rate remains at approximately $92\%$, adding 15,056 additional people into the ED will require the construction of almost 6,000 new dwellings. Conversely, Derrylaur in Co. Galway's projected decline to an average of just 9 inhabitants would mean that even if no more dwellings were constructed in the ED during the simulation period and all of the inhabitants lived alone, the occupancy rate would still only be $16.6\%$. Of course, this availability of homes could make Derrylaur an attractive place to move to, and this is something that could be considered in future versions of the simulation.

With regards to education, although the overall population is projected to increase significantly, the number of children in primary school is expected to fall approximately $22\%$. This is because of the ageing population and declining birth rates. Without the construction or demolition of any primary schools, the mean number of students per school would drop from approximately $183$ to approximately $142$. This could significantly ease the pressure on teachers who have demanded Ireland reduces its class size closer to the EU average \citep{into2020classSize}. The ageing population will also have an impact on public expenditure. In 2025, Ireland has budgeted €8.08 billion towards state pension payments with retirees comprising about 12\% of the overall population. Assuming that the amount spent per capita remains constant (and not adjusting for inflation), the 2057 retiree population percentage of $\approx25\%$ will necessitate spending of almost €17 billion. This combined with the projection that the number of widow(er)s is expected to increase by approximately $34\%$ (which will affect the widow(er)'s pension spending), means that the government may be required to spend almost €20 billion on pension payments in 2057.

Future extensions of the model may include additional characteristics such as health status, ethnicity, or Irish-speaking ability. Another possible extension would be to group individuals into households which would necessitate the implementation of modules such as cohabitation and children leaving home. More in-depth analysis of a specific characteristic, like education, could also provide valuable insights in that domain. Nonetheless, the current model provides a strong foundation for analysing demographic change and informing evidence-based planning in Ireland.

\section*{Acknowledgements}
This work was funded by the University of Galway College of Science and Engineering Postgraduate Scholarship.

\section*{Declaration of Competing Interests}
The authors declare that they have no known competing financial interests or personal relationships that could have appeared to influence the work reported in this paper.

\bibliography{references}

\appendix
\section*{Appendices}
\renewcommand{\thesection}{A.\arabic{section}}

\section{Repository Link}
\label{sec:link}
The microsimulation code as well as the resulting populations can be found at \url{https://github.com/SCC-git/ireland-microsim}. Note that some statistics are calculated from the LFS microdata, which is only available to approved applicants at \url{https://www.ucd.ie/issda/accessdata/issdadatasets/}.

\section{Dependency Ratios}
\label{sec:dependency_ratios}
\input{m2_population_comparison_table}

\input{m3_population_comparison_table}

Tables \ref{tab:comparisonTableM2} and \ref{tab:comparisonTableM3} display the regional young and old dependency ratios for the M2 and M3 scenarios, respectively. As in the M1 case, an acceptable goodness-of-fit is evident given the various possible differences in the implementation of fertility, mortality, migration, etc., between the microsimulation and DCM.

\clearpage

\section{Assumptions}
\label{sec:assumptions}

\begin{longtable}{|p{0.4\linewidth}|p{0.25\linewidth}|P{0.25\linewidth}|}

    \hline
    \textbf{Statistic} & \textbf{Based on} & \textbf{Source}  \\
    \hline
    Regional age-group-specific fertility rates & 2022 values & \url{https://data.cso.ie/table/VSA104} \\ \hline
    Single year-of-age within 5-year age-group proportions (national level) & 2022 values & \url{https://data.cso.ie/table/F1002} \\ \hline
    Proportions of students within individual further and higher education courses (all students in Ireland) & Proportions of non-students with qualifications in 2022 & \url{https://data.cso.ie/table/EDQ01} \\ \hline
    Proportions of young students studying for specific third-level qualifications (national level) & Proportions of graduates with qualifications in 2019 & \url{https://data.cso.ie/table/HEO01} \\ \hline
    Proportions of older students studying for specific qualifications (national level) & Proportions of older adults with qualifications in 2022 synthetic population & Synthetic Population \\ \hline
    Same-sex proportion of marriages (national level) & Number of same- and opposite- sex marriages registered in 2024 & \url{https://data.cso.ie/table/VSB02} \& \url{https://data.cso.ie/table/VSA44}  \\ \hline
    Age- and sex-specific mortality rates (national level) & 2016 life tables & \url{https://data.cso.ie/table/VSA32} \\ \hline
    County-to-county migrant counts & 2022 or 2016 values (based on scenario) & \url{https://data.cso.ie/table/F1024} \\ \hline
    Within-county migrant counts (county level) & 2022 values & \url{https://data.cso.ie/table/SAP2022T2T3CTY} \\ \hline
    EDs' proportion of their county's within-county migrants & 2022 values & \url{https://data.cso.ie/table/SAP2022T2T3ED} \\ \hline
    Proportion of county's within-county migrants by age-and-sex bracket (ED level) & 2022 values & \url{https://data.cso.ie/table/F1038} \\ \hline
    Proportion of county's different-county immigrants by age-and-sex bracket (ED level) & 2022 values & \url{https://data.cso.ie/table/F1038} \\ \hline
    Regional proportions of Ireland's emigrants & 2011-2016 values & \url{https://www.cso.ie/en/releasesand publications/ep/p-rpp/regionalpopulati onprojections2017-2036/} \\ \hline
    Proportions of international emigrants in sex-and-age-group brackets (national level) & 2017 to 2022 average values & \url{https://data.cso.ie/table/PEA03} \\ \hline
    Proportions of international immigrants in sex-and-age-group brackets (national level) & 2017 to 2022 average values & \url{https://data.cso.ie/table/PEA03} \\ \hline
    EDs' proportions of Ireland's total international immigrants & 2022 values & \url{https://data.cso.ie/table/SAP2022T2T3ED} \\ \hline
    Age-group specific within-marriage proportions of births (national level) & 2022Q4 values & \url{https://data.cso.ie/table/VSQe77} \\ \hline
    Separations as a proportion of the number of married people (national level) & 2022 values for separation and married people & Separations: \url{https://www.courts.ie/content/annual-report-2022-published}, Married population: \url{https://data.cso.ie/table/F3001} \\ \hline
    Marriage Rate (national level) & 2023 value & \url{https://www.cso.ie/en/releasesand publications/ep/p-mar/marriages2023/mainresults/} \\ \hline
    Parents' broad education to child's education (national level) & 2023 values & \url{https://data.cso.ie/table/SID23} \\ \hline
    Early school leavers percentage (national level) & 2004-2024 values & \url{https://data.cso.ie/table/EDQ07} \\ \hline
    Dropout rates at all NFQ levels (national level) & 2022 values & \url{https://hea.ie/statistics/data-for-download-and-visualisations/students/completion/completion-analysis-200809-200910-201011-entrants/} \\ \hline
    Outcomes for dropouts (national level) & 2022 values & \url{https://data.cso.ie/table/NPO03} \\ \hline
    Enrolment rates by previous education (national level) & 2022 values & Further Education: \url{https://data.cso.ie/table/FEO08}, Higher Education: \url{https://hea.ie/statistics/graduate-outcomes-data-and-reports/graduate-outcomes-all-years-2017-2024/} \\ \hline
    Outcomes for graduates (national level) & 2022 values & \url{https://hea.ie/statistics/graduate-outcomes-data-and-reports/graduate-outcomes-all-years-2018-2023/} \\ \hline
    Regional proportions of adults who are students by age group and sex & 2022 values & \url{https://data.cso.ie/table/F7013} \\ \hline
    Proportion of young adults that are students (national level) & 2022 value & \url{https://data.cso.ie/table/F8050} \\ \hline
    Primary economic status by age group, sex, citizenship, and highest level of education attained & 2023Q3 LFS Data & LFS \\ \hline    
    \caption{All of the explicit historic-data-based assumptions made in the microsimulation and their sources. All data is specific to Ireland. Statistics are arranged based on their appearance order in the code. Where the level of granularity of the data is not obvious, it is specified in brackets after the statistic.}
    \label{tab:assumptions_and_sources}
\end{longtable}


\end{document}

%% file: regions_and_counties.tex
\begin{table}[!htb]
    \centering
    \begin{tabular}{|c|c|}
    \hline
    Region & Counties \\
    \hline
    \multirow{5}{*}{Border} & Donegal \\ 
    & Sligo  \\ 
    & Leitrim  \\ 
    & Cavan  \\ 
    & Monaghan  \\ 
    \hline
    \multirow{3}{*}{West} & Galway \\ 
    & Mayo  \\ 
    & Roscommon  \\ 
    \hline
    \multirow{3}{*}{Mid-West} & Clare \\ 
    & Tipperary  \\ 
    & Limerick  \\ 
    \hline
    \multirow{4}{*}{South-East} & Waterford \\ 
    & Kilkenny  \\ 
    & Carlow \\
    & Wexford  \\ 
    \hline
    \multirow{2}{*}{South-West} & Cork \\ 
    & Kerry  \\ 
    \hline
    \multirow{1}{*}{Dublin} & Dublin \\ 
    \hline
    \multirow{4}{*}{Mid-East} & Wicklow \\ 
    & Kildare  \\ 
    & Meath  \\ 
    & Louth  \\ 
    \hline
    \multirow{4}{*}{Midland} & Longford \\ 
    & Westmeath  \\ 
    & Offaly  \\ 
    & Laois  \\ 
    \hline
    \end{tabular}
    \caption{Ireland's NUTS3 Regions and their constituent counties}
    \label{tab:regionsCounties}
\end{table}

%% file: m1_population_comparison_table.tex
\begin{table}[!htb]
    \centering
    \begin{tabular}{c|c|c|c|c}
    Region & Statistic & CSO & SEMIPro & Abs. Diff. \\
    \hline
    \multirow{2}{*}{Border} & YDR & 23.2 & 23.5 & 1.1\% \\
    & ODR & 37.5 & 38.0 & 1.3\% \\ 
    \hline
    \multirow{2}{*}{Mid-West} & YDR & 22.4 & 22.7 & 1.1\% \\
    & ODR & 37.6 & 38.6 & 2.5\% \\
    \hline
    \multirow{2}{*}{South-West} & YDR & 20.7 & 20.3 & 1.8\% \\
    & ODR & 36.5 & 36.7 & 0.6\% \\
    \hline
    \multirow{2}{*}{West} & YDR & 21.9 & 21.9 & 0.0\% \\
    & ODR & 37.2 & 37.4 & 0.7\% \\
    \hline
    \multirow{2}{*}{Midland} & YDR & 22.2 & 22.9 & 3.1\% \\
    & ODR & 34.4 & 35.4 & 2.8\% \\
    \hline
    \multirow{2}{*}{Mid-East} & YDR & 22.5 & 22.8 & 1.3\% \\
    & ODR & 32.4 & 33.4 & 3.1\% \\
    \hline
    \multirow{2}{*}{South-East} & YDR & 22.6 & 22.8 & 0.9\% \\
    & ODR & 35.4 & 37.7 & 6.1\% \\
    \hline
    \multirow{2}{*}{Dublin} & YDR & 19.0 & 17.7 & 7.5\% \\
    & ODR & 30.5 & 29.2 & 4.4\% \\
    \end{tabular}
    \caption{Comparison of population size, young dependency ratio (YDR) and old dependency ratio (ODR) for the M1 international migration scenario.}
    \label{tab:comparisonTableM1}
\end{table}

%% file: m2_population_comparison_table.tex
\begin{table}[!htb]
    \centering
    \begin{tabular}{c|c|c|c|c}
    Region & Statistic & CSO & SEMIPro & Abs. Diff. \\
    \hline
    \multirow{2}{*}{Border} & YDR & 23.2 & 23.5 & 1.2\% \\
    & ODR & 39.0 & 39.7 & 1.7\% \\
    \multirow{2}{*}{Mid-West} & YDR & 22.2 & 22.7 & 2.4\% \\
    & ODR & 38.9 & 40.1 & 3.1\% \\
    \multirow{2}{*}{South-West} & YDR & 20.5 & 20.2 & 1.3\% \\
    & ODR & 37.8 & 38.2 & 1.1\% \\
    \multirow{2}{*}{West} & YDR & 21.7 & 21.8 & 0.6\% \\
    & ODR & 38.5 & 38.9 & 0.9\% \\
    \multirow{2}{*}{Midland} & YDR & 21.9 & 23.0 & 4.8\% \\
    & ODR & 35.6 & 36.6 & 2.7\% \\
    \multirow{2}{*}{Mid-East} & YDR & 22.4 & 22.8 & 1.6\% \\
    & ODR & 33.3 & 34.4 & 3.3\% \\
    \multirow{2}{*}{South-East} & YDR & 22.5 & 22.7 & 0.8\% \\
    & ODR & 36.7 & 39.0 & 6.0\% \\
    \multirow{2}{*}{Dublin} & YDR & 18.7 & 17.5 & 6.7\% \\
    & ODR & 32.1 & 30.9 & 3.8\% \\
    \end{tabular}
    \caption{Comparison of young dependency ratio (YDR) and old dependency ratio (ODR) for the M2 international migration scenario.}
    \label{tab:comparisonTableM2}
\end{table}

%% file: m3_population_comparison_table.tex
\begin{table}[!htb]
    \centering
    \begin{tabular}{c|c|c|c|c}
    Region & Statistic & CSO & SEMIPro & Diff. \\
    \hline
    \multirow{2}{*}{Border} & YDR & 23.1 & 23.7 & 2.4\% \\
& ODR & 41.7 & 42.5 & 1.9\% \\
\multirow{2}{*}{Mid-West} & YDR & 22.2 & 22.9 & 3.1\% \\
& ODR & 40.8 & 42.2 & 3.2\% \\
\multirow{2}{*}{South-West} & YDR & 20.4 & 20.3 & 0.6\% \\
& ODR & 39.8 & 40.4 & 1.4\% \\
\multirow{2}{*}{West} & YDR & 21.8 & 21.9 & 0.6\% \\
& ODR & 40.4 & 40.9 & 1.3\% \\
\multirow{2}{*}{Midland} & YDR & 22.1 & 23.2 & 4.9\% \\
& ODR & 37.7 & 38.6 & 2.4\% \\
\multirow{2}{*}{Mid-East} & YDR & 22.4 & 22.9 & 2.1\% \\
& ODR & 34.4 & 35.7 & 3.7\% \\
\multirow{2}{*}{South-East} & YDR & 22.5 & 22.8 & 1.2\% \\
& ODR & 37.9 & 40.7 & 6.9\% \\
\multirow{2}{*}{Dublin} & YDR & 18.7 & 17.5 & 7.1\% \\
& ODR & 34.8 & 33.4 & 4.3\% \\
    \end{tabular}
    \caption{Comparison of young dependency ratio (YDR) and old dependency ratio (ODR) for the M3 international migration scenario.}
    \label{tab:comparisonTableM3}
\end{table}